\documentclass{article}

\usepackage{arxiv} 
\usepackage{authblk}

\usepackage{amsfonts}
\usepackage{algorithm, algpseudocode}
\usepackage{multirow, multicol, makecell, array}
\usepackage{subcaption}
\usepackage{enumitem}
\usepackage{comment}
\usepackage{url}
\usepackage{xcolor}
\usepackage{graphicx}
\usepackage[pdftex]{hyperref}
\usepackage{amsmath, nccmath}
\usepackage{bm}

\newcommand\tab[1][0.5cm]{\hspace*{#1}}
\algnewcommand{\LineComment}[1]{\State \(\triangleright\) #1}
\algnewcommand{\LineCommentX}[1]{\Statex \(\triangleright\) #1}

\let\oldtabular\tabular 
\renewcommand{\tabular}{\footnotesize\oldtabular}

\algtext*{EndWhile}
\algtext*{EndFor}
\algtext*{EndIf}

\newbox{\orcid}\sbox{\orcid}{\includegraphics[scale=0.06]{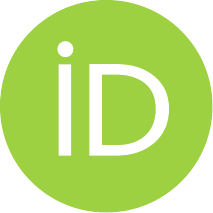}}

\author[1,2,4]{
	\href{https://orcid.org/0000-0002-3484-2535}
	{\usebox{\orcid}\hspace{1mm}Yesdaulet Izenov}
}
\author[1,3]{
	\href{https://orcid.org/0009-0001-7161-8444}
	{\usebox{\orcid}\hspace{1mm}Asoke Datta}
}
\author[1,3]{
	\href{https://orcid.org/0000-0001-8407-7563}
	{\usebox{\orcid}\hspace{1mm}Brian Tsan}
}
\author[1,3]{
	\href{https://orcid.org/0009-0000-3276-6762}
	{\usebox{\orcid}\hspace{1mm}Abylay Amanbayev}
}
\author[1,3]{
	\href{https://orcid.org/0000-0002-7018-9043}
	{\usebox{\orcid}\hspace{1mm}Florin Rusu}
}

\affil[1]{University of California Merced, Merced, California, USA}
\affil[2]{Nazarbayev University, Astana, Kazakhstan}
\affil[3]{\texttt{\{adatta2, btsan, amanbayev, frusu\}@ucmerced.edu}}
\affil[4]{\texttt{yesdaulet.izenov@nu.edu.kz}}

\title{Spanning Tree-based Query Plan Enumeration}

\renewcommand{\headeright}{Izenov et al.}
\renewcommand{\undertitle}{Technical Report}
\renewcommand{\shorttitle}{Spanning Tree-based Query Plan Enumeration}

\hypersetup{
    pdftitle = {Spanning Tree-based Query Plan Enumeration},
    pdfauthor = {Yesdaulet Izenov, Asoke Datta, Brian Tsan, Abylay Amanbayev, Florin Rusu},
    pdfkeywords = {query optimization, query planning, query plan enumeration, join ordering, spanning tree}
}

\begin{document}
\maketitle

\begin{abstract}
Query optimizer architectures typically employ either an exhaustive or heuristic strategy to enumerate query execution plans. Exhaustive strategies enumerate every possible query plan to ensure plan optimality. However, exhaustive strategies become computationally expensive to enumerate a massive number of execution plans for queries with a large number of joins. Conversely, heuristic strategies are designed to enumerate a single query plan. However, while selecting a query plan in a polynomial time, heuristic strategies may select significantly suboptimal query plans. In the case of large queries, shifting from an exhaustive to a heuristic strategy can result in a notable decline in the quality of query plans, subsequently leading to less robust query optimizer performance. An even worse scenario is to reimplement the optimizer to a different enumeration strategy causing significant expenses in terms of development.

In this work, we define the problem of finding an optimal query plan as finding spanning trees with low costs. This approach empowers the utilization of a series of spanning tree algorithms, thereby enabling systematic exploration of the plan search space over a join graph. Capitalizing on the polynomial time complexity of spanning tree algorithms, we present ESTE -- Ensemble Spanning Tree Enumeration strategy. In this work, we employ two conventional spanning tree algorithms, Prim's and Kruskal's, together to enhance the robustness of the query optimizer. In ESTE, multiple query plans are enumerated exploring different areas of the search space. This positions ESTE as an intermediate strategy between exhaustive and heuristic enumeration strategies. We show that ESTE is more robust in identifying efficient query plans for large queries. In the case of data change and workload demand increase, we believe our approach can be a cheaper alternative to maintain optimizer robustness by integrating additional spanning tree algorithms rather than completely changing the optimizer to another plan enumeration algorithm. Experimental evaluations show ESTE offers better consistency in plan quality and optimization time devoting minimal extra time to optimization.
\end{abstract}


\section{INTRODUCTION}\label{sec:intro}
Plan enumeration is a critical component of query optimization. To find an optimal query execution plan, conventional query optimizer architectures employ either an exhaustive or heuristic plan enumeration strategy. In the case of large queries, some architectures offer flexibility by switching from exhaustive to heuristic strategy~\cite{postgres,Neumann:AOVLJQ:sigmod-2018}. This implies a trade-off between plan optimality and optimization time. Exhaustive strategy requires enumerating every possible query plan to ensure the selection of the optimal one, yet this thoroughness results in considerable optimization time. On the other hand, heuristic strategies enumerate a single plan which may not be optimal but offers faster optimization time. Shifting between these two extremes, exhaustive and heuristic strategies, can lead to a significant change -- increase or decrease in optimizer performance. Particularly, deterministic heuristic strategies, such as greedy algorithms, tend to get trapped in local optima -- consistently yielding suboptimal plans for a query. As an intermediate strategy, plan enumeration can enumerate more than a single query plan by introducing randomness to diversify the set of enumerated query plans~\cite{Steinbrunn:HRO:vldb-1997,Ioannidis:RANDOM:sigmod-1990,Swami:LJQ:sigmod-1989,postgres,Swami:IIO:sigmod-1988,Ioannidis:ANNEAL:sigmod-1987}. This increases the chances to locate superior query plans thus reducing the gap between plan optimality and optimization time. While randomized heuristic strategies consume an additional optimization time, this strategy ensures its feasibility for practical use in large queries. However, the optimizer performance becomes less predictable and interpretable due to the random behavior in the search space exploration.

In this paper, we define query optimization as the problem of finding spanning trees with low costs. The entire search for a query plan is defined in terms of join graph edges. Thus the objective is to identify an ordered sequence of edges that span all vertices in the join graph while minimizing the total sum of edge weights. Unlike in MVP and GreedyJoinOrdering-3~\cite{Lee:MVP:tkde-2001,Moerkotte:BQO-book:2023}, unnecessary joins and cross-joins are naturally avoided from query plans by representing tables as vertices and operating over existing edges in the join graph.  Formulation over graph edges enables extensive application of spanning tree algorithms in the domain of query optimization. Building on this foundation, we introduce Ensemble Spanning Tree Enumeration (ESTE) as a novel, intermediate strategy for plan enumeration. Unlike in GOO~\cite{Fegaras:NEWHEU:dexa-1998} and GreedyJoinOrdering-3~\cite{Moerkotte:BQO-book:2023}, ESTE utilizes a set of fast, spanning tree-based algorithms to enumerate multiple spanning trees with low costs, each algorithm exploring different areas of the search space over a join graph. Unlike in directed plan graph representations~\cite{Negi:FLOWLOSS:pvldb-2021,Haffner:HEU:sigmod-2023}, the search space does not increase due to its simple query representation. Within ESTE, we employ two classical spanning tree algorithms, Prim's and Kruskal's, to enhance the robustness of the query optimizer. Unlike in IK-KBZ and GreedyJoinOrdering-3~\cite{Krishnamurthy:KBZ:vldb-1986,Moerkotte:BQO-book:2023}, the resulting spanning trees are not restricted to linear query plans. Furthermore, unlike in previous work~\cite{Krishnamurthy:KBZ:vldb-1986,Fegaras:NEWHEU:dexa-1998,Moerkotte:BQO-book:2023}, we utilize a cost function~\cite{Leis:JOB:vldb-2018,Leis:QOREALLY:pvldb-2015} that considers physical operators such as multiple join algorithms and scan operators. Thereby the resulting spanning trees are physical plans rather than logical plans. The flexibility of ESTE allows for the incorporation of additional spanning tree algorithms in response to changes in data and workload, thus maintaining robustness and even enhancing the optimizer performance. We believe that our approach offers a more cost-effective way to sustain optimizer robustness, favoring the integration of new algorithms over a complete change and redevelopment of the optimizer for different plan enumeration strategies. We outline the key technical contributions as follows:

\begin{itemize}[leftmargin=*,noitemsep,nolistsep,topsep=0pt]
  	\item We present a novel perspective on query optimization by framing it as a problem of finding spanning trees with low costs.
  	\item We introduce Ensemble Spanning Tree Enumeration (ESTE), a novel approach that systematically harnesses unique plan search space exploration methods inherent to Prim's and Kruskal's algorithms. Thus multiple query plans are enumerated to provide robustness to query optimization by employing a series of spanning tree algorithms over the join graph.
  	\item In the scope of this paper, we adapt both Prim's and Kruskal's algorithms to account for changes in edge weights as spanning trees are built. The resulting spanning trees are physical plans, with various binary tree shapes, rather than logical plans.
  	\item We evaluate query plans produced by ESTE and compare to exhaustive enumeration and GOO~\cite{Fegaras:NEWHEU:dexa-1998} which is recognized as one of top-performing heuristic methods~\cite{Leis:JOB:vldb-2018,Leis:QOREALLY:pvldb-2015,Neumann:AOVLJQ:sigmod-2018,Haffner:HEU:sigmod-2023}. We also examine ESTE performance across chain, cycle, start and clique graph topologies and compare ESTE to other well-known and recently proposed heuristic methods.
	\item We examine the performance behavior of ESTE in the presence of cardinality estimation errors. As expected, we observe a decline in efficiency across all enumeration methods, including exhaustive enumeration. The results exhibit a higher overall performance of ESTE compared to the other considered enumeration methods, while also maintaining a low optimization time.
\end{itemize}
\section{SPANNING TREE-BASED QUERY OPTIMIZATION}\label{sec:spanning_tree}
In this section, we define classical cost-based query optimization as finding spanning trees with low costs over a join graph. The entire search procedure -- including four elements, namely cost model, cardinality estimation, plan search space, and plan enumeration -- are described in terms of join graph edges.

\begin{figure}[htbp]
	\centering
	\includegraphics[scale=0.61]{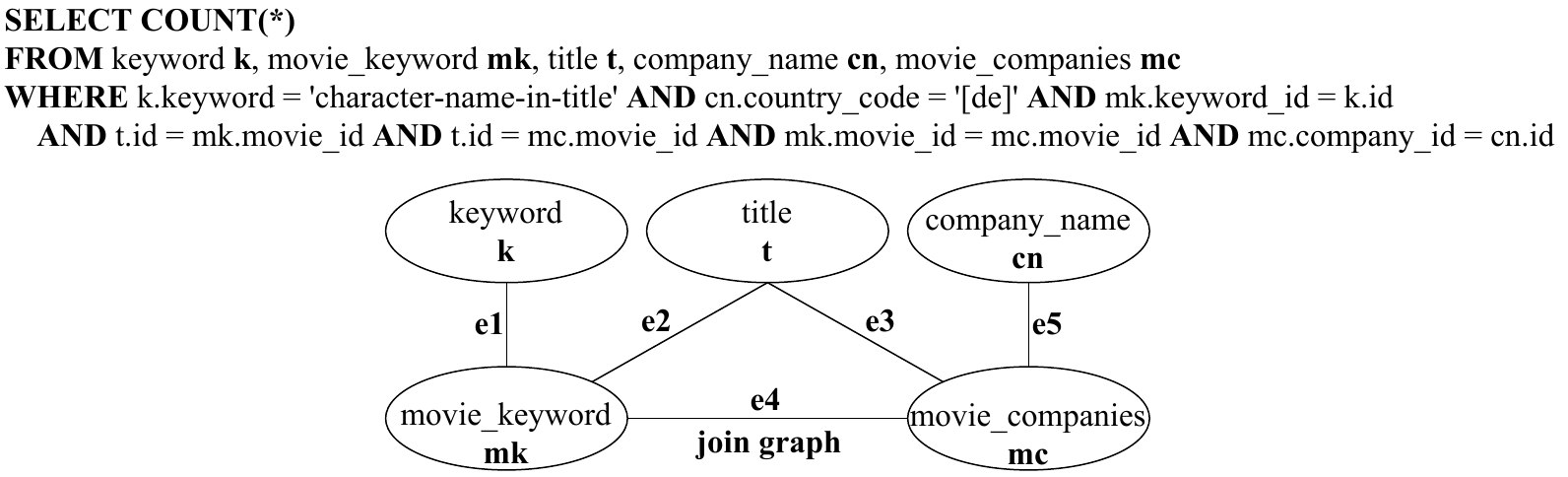}
	\caption{SQL statement for JOB query 2a and its join graph.}
	\label{fig:2a_query}
\end{figure}

\textbf{Join graph.}
In Figure~\ref{fig:2a_query}, the SQL statement has 2 point selection predicates and 5 join predicates connecting 5 tables. Further, the query is represented as an undirected join graph $G(V, E)$ in which every table is a vertex $v \in V$ connected with edges $e \in E$ representing the equi-join predicates. Vertices form a single component connected with a list of weighted edges $E \subseteq \{(v1, v2, w) \mid v1, v2 \in V, w \in \mathbb{R^+}\}$. For example, the join predicate \textit{mk.keyword\_id = k.id} is the edge $e1$ which connects the vertices \textit{mk} and \textit{k} in the join graph. The edges $e2$, $e3$, and $e4$ form the cycle between \textit{mk}, \textit{t}, and \textit{mc}. In the case of no cycles in a join graph, the number of edges is $|E| = |V| - 1$. This is also the minimum number of edges required to have a connected graph. Non-existent edges $e \notin E$ in the join graph are not considered. The topology of a join graph can be classified as either a chain, cycle, star, or clique. The number of edges increases from $|V| - 1$ for chain and star to $(|V| \cdot |V - 1|) / 2$ for clique while the cyclic join graph falls between these two extremes.

\textbf{Query plan.}
A query plan $\mathcal{P}$ is defined as a spanning tree -- acyclic subgraph -- of the join graph $G$. A spanning tree is an ordered sequence of $|V|-1$ edges that spans all $V$ vertices in $G$, thus forming a single connected component. Cross-joins between tables are naturally avoided since a spanning tree is built over existing edges $e \in E$ in $G$. The order of edges in a spanning tree corresponds to a path which corresponds to the join order of the tables in the query plan. For instance, in Figure~\ref{fig:2a_query}, the edge sequence $(e1, e2, e3, e5)$ corresponds to spanning tree $\mathcal{P} = \{(\textit{mk}-\textit{k}), (\textit{t}-\textit{mk}), (\textit{t}-\textit{mc}), (\textit{mc}-\textit{cn})\}$ which results into join order $(mk \Join k \Join t \Join mc \Join cn)$. While a spanning tree does not include cycles, this does not mean that the joins corresponding to the edges not included in the spanning tree are dropped from the query plan. Instead, these join predicates are kept and treated as filters. However, in the case of transitive join predicates, these joins can be eliminated from the query plan. Concretely, the example spanning tree $\mathcal{P}$ does not include the edge $e4$ which is the join between $(\textit{mk}-\textit{mc})$ corresponding to the join predicate \textit{mk.movie\_id = mc.movie\_id}. This predicate can be evaluated together with the join $(\textit{t}-\textit{mc})$ since \textit{mk} and \textit{mc} are both part of current connected component $\{(\textit{mk}-\textit{k}), (\textit{t}-\textit{mk}), (\textit{t}-\textit{mc})\}$. However, this is not necessary for this query because the join cycle is based on transitive predicates -- which already implies the ignored join condition. This simplification does not apply to general cyclic queries such as those considered by worst-case optimal joins~\cite{Ngo:WORST:pods-2012}.

\textbf{Query subplan.}
A query subplan is defined as a query plan over a subgraph of the join graph. For instance, $\mathcal{P}_{mk \Join k \Join t} = \{(\textit{mk}-\textit{k}), (\textit{t}-\textit{mk})\}$ corresponds to the query subplan over tables \textit{k}, \textit{mk}, and \textit{t}, and its associated subgraph.

\begin{figure*}[htbp]
	\centering
	\includegraphics[width=\textwidth]{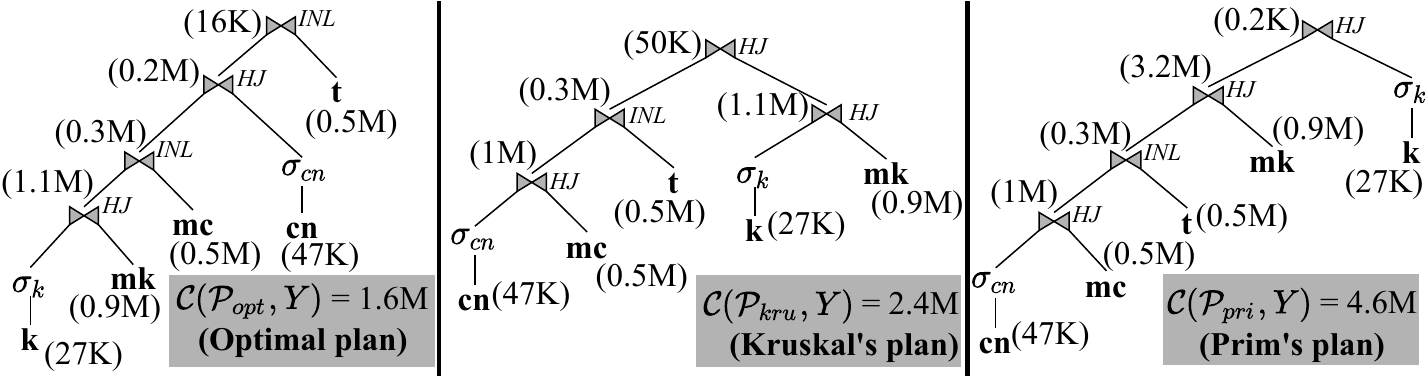}
 	\caption{Three query plans ($\mathcal{P}_{\textit{opt}}$, $\mathcal{P}_{\textit{kru}}$ and $\mathcal{P}_{\textit{pri}}$) for query 2a selected using 3 different query plan enumeration algorithms (Exhaustive, Kruskal, and Prim) along with their costs $\mathcal{C}$ computed using exact cardinalities $Y$.}
 	\label{fig:2a_query_plans}
\end{figure*}

\textbf{Linear query plan.}
The sequence, in which edges are chosen, directly influences the resulting shape of the spanning tree. Given that an edge inherently involves a connection between two vertices in the join graph, the shape of the corresponding spanning tree invariably adheres to a binary shape. The leftmost and rightmost spanning trees depicted in Figure~\ref{fig:2a_query_plans} exhibit linear query plans. In building a linear query plan, the invariant is to keep a single acyclic connected component and incrementally add adjacent edges until all the vertices are covered. At every join, either of the vertices is set to be a base table.

\textbf{Bushy query plan.}
Bushy spanning trees exhibit a distinctive structure that sets them apart from strictly linear spanning trees. A defining attribute of bushy spanning trees is the possibility for an internal vertex to have both of its vertices as internal vertices. The bushy spanning tree depicted in the middle of Figure~\ref{fig:2a_query_plans} has the root vertex which has two subtrees, $\mathcal{P}_{cn \Join mc \Join t} = \{(\textit{mc}-\textit{cn}), (\textit{t}-\textit{mc})\}$ and $\mathcal{P}_{mk \Join k} = \{(\textit{mk}-\textit{k})\}$, as its vertices each forming two separate acyclic connected components. Bushy spanning trees do not follow a strictly linear pattern in their order, thus they can be effectively capitalized on opportunities for parallelism and flexible execution ordering to optimize query performance.

\subsection{Cost Model}\label{subsec:cost-model}
To evaluate and compare spanning trees derived from a join graph, a quantifiable cost is assigned to each spanning tree using a predefined analytical cost function. The optimal query plan is the minimum spanning tree, as it is anticipated to have the fastest execution time. However, formulating a cost function that can accurately reflect execution time is challenging. While disk-based cost functions remain crucial, a main memory cost function can provide an accurate estimate of the actual query execution time in modern database systems. Virtually all main memory cost functions are defined in terms of the number of tuples processed by the physical operators in a query plan, while disk-based cost functions consider block reads instead of tuples~\cite{Negi:FLOWLOSS:pvldb-2021,Wolf:ROBUST:pvldb-2018}. Leis et al.~\cite{Leis:JOB:vldb-2018,Leis:QOREALLY:pvldb-2015} proposed a simple main memory cost function $\mathcal{C}$ that incorporates physical operators shown in Equation~\ref{eq:cost-model}. Cost function $\mathcal{C}$ sums the costs of internal vertices while the costs of leaf vertices are internally used to compute the cost of internal vertices. In the leaf vertices, the size of a base table $R$ is multiplied by $\tau = 0.2$ to differentiate a base table scan from the intermediate join cost. Notice a base table with selection predicates $\sigma$ still requires a full table scan. For an internal vertex, the cost function considers hash join $\Join^{HJ}$ and index nested-loop join $\Join^{INL}$. In this work, we select a vertex with the smallest cost to build a hash table. To differentiate between hash lookup and index lookup, $\lambda = 2$ is used assuming the existence of primary and foreign key indexes on all join attributes of base tables -- otherwise, the hash join is utilized.

\begin{equation}
	\mathcal{C}(\mathcal{P}) = 
	\begin{cases}
		\tau \times |R| & \textbf{if}~\; \mathcal{P} = R \lor \mathcal{P} = \sigma(R) \\
		|\mathcal{P}| + |\mathcal{P}_1| + \mathcal{C}(\mathcal{P}_1) + \mathcal{C}(\mathcal{P}_2) & \textbf{if}~\; \mathcal{P} = \mathcal{P}_1 \Join^{HJ} \mathcal{P}_2 \\
		\mathcal{C}(\mathcal{P}_1) + \lambda \times |\mathcal{P}_1| \times \max(\frac{|\mathcal{P}_1 \Join R|}{|\mathcal{P}_1|}, 1) & \textbf{if}~\; \mathcal{P} = \mathcal{P}_1 \Join^{INL} \mathcal{P}_2 \land (\mathcal{P}_2 = R \lor \mathcal{P}_2 = \sigma(R)) \\
	\end{cases}
	\label{eq:cost-model}
\end{equation}

Cost function $\mathcal{C}$ dictates whether a spanning tree has a left-deep, right-deep or zig-zag structure in linear query plans. Left-deep spanning tree sets the right vertex to be a base table as a leaf vertex while a right-deep spanning tree sets the left vertex to be a base table. A zig-zag spanning tree establishes a balance by alternating the roles of the left and right vertices at each join level. Depending on the costs, leftmost and rightmost spanning trees, depicted in Figure~\ref{fig:2a_query_plans}, can have left-deep, right-deep or zig-zag structures. The same concept is applied to linear subtrees of bushy spanning trees. The costs of three spanning trees $\mathcal{P}_{\textit{opt}}$, $\mathcal{P}_{\textit{kru}}$ and $\mathcal{P}_{\textit{pri}}$ are displayed in Figure~\ref{fig:2a_query_plans}. For example, in the case of the optimal plan $\mathcal{P}_{\textit{opt}}$, the cost $\mathcal{C}(\mathcal{P}_{\textit{opt}}, Y) = 1.1M + 0.3M + 0.2M + 16K = 1.6M$ is the sum of the costs corresponding to its 2-way join $(mk \Join k)$, 3-way join $(mk \Join k \Join mc)$, 4-way join $(mk \Join k \Join mc \Join cn)$, and 5-way join $(mk \Join k \Join mc \Join cn \Join t)$, respectively. It is important to note that the cost of $(mk \Join k \Join mc)$ is not the cost of the corresponding edge $(\textit{mk}-\textit{mc})$ from the join graph which is $39M$. This is because only the tuples in the 2-way join $mk \Join k$ are subsequently joined with $\textit{mc}$ -- not the entire $\textit{mk}$.

It is critical to employ accurate cost functions in both main memory and disk-based query processing, however, both can significantly suffer from underlying poor cardinality estimates. This, in turn, affects the join order which can significantly influence the execution time -- inefficient ordering of table joins can result in unnecessary data access.

\subsection{Cardinality Estimation}
In Equation~\ref{eq:cost-model}, the cost function $\mathcal{C}$ includes $\mathcal{P}$, $\mathcal{P}_1$, and $R$ as cardinalities of input tables to compute the costs of physical operators. However, exact cardinality $Y$ can be computed exactly only by executing the join. This is a paradoxical situation in query optimization. The query optimizer is supposed to determine the optimal plan in which order to perform the joins without actually performing them. This is where cardinality estimation comes into the picture. The role of cardinality estimation is to ``guess'' the exact cardinality $Y$ without executing the join. In PostgreSQL~\cite{postgres}, cardinality estimates are computed based on histograms, most frequent values, and distinct element statistics on the base tables. Any join estimate is computed by combining these statistics into simple arithmetic formulas that make general assumptions on uniformity, inclusion, and independence~\cite{Lohman:QOSP:2014}. Consequently, cardinality estimations $\hat{Y}$ are fed into the cost function $\mathcal{C}$ in lieu of exact cardinalities, allowing the query optimizer to compute the cost of a query plan. Henceforth, two query plans can be ranked according to their estimated cost -- instead of the exact cost. As long as the ranks of the query plans based on the exact and estimated cost are identical, estimates can be a direct replacement for the exact cardinalities.

From a statistical perspective, the goal of cardinality estimation is to minimize the difference between $Y$ and $\hat{Y}$. However, from the cost model perspective, this is not necessary. Instead, what is required is that the costs of two different query plans satisfy the same query plan when computed based on estimates $\hat{Y}$ and when using the exact cardinalities $Y$. While accurate estimates imply the optimal query plan, accurate estimates are not required as long as they have similar errors. For example, estimates that are $100\times$ larger than the exact value are extremely inaccurate. However, the cost model treats them identically and results in two costs that have the same query plan. Thus, estimation does not impact the cost model negatively. Obtaining accurate estimates is a challenging task. Despite a rich number of cardinality estimation techniques~\cite{Leis:index-join-sample:cidr-2017,Muller:ISECKSS:pvldb-2018,Izenov:COMPASS:sigmod-2021,Cai:PCETUB:sigmod-2019,Ioannidis:PESJR:sigmodrec-1991,Poosala:SEW:vldb-1997,Kiefer:KDE:pvldb-2017,Malik:BBAQCE:cidr-2007,Liu:CEUNN:cascon-2015,Kipf:LCECJDL:cidr-2019,Kipf:ECDL:arxiv-2019,Woltmann:CEL:aiDM-2019,Ortiz:EADLCE:arxiv-2019}, cardinality estimation error is inevitable triggering a chain of more errors in subsequent optimizer components. These challenges underline the importance of ongoing research into improving the accuracy of cardinality estimates to enhance the efficiency of query optimization.

\subsection{Search Space}\label{subsec:search-space}
The search space is a set of every possible query plan in which all tables are joined~\cite{Leis:JOB:vldb-2018}. Two query plans of the same shape but with different join orders are two different query plans. We first define the search space in terms of the number of binary trees using vertices $V$ of a join graph $G$. Further, we define the search space in terms of the number of spanning trees based on edges $E$ of a join graph $G$. We show that operating on edges incorporates connectivity information inherent in the join graph which naturally avoids query plans involving cross-joins.

\textbf{Search space in terms of graph vertices.}
The search space is typically defined as a set of distinct binary tree shapes in which each vertex can be any of the $V$ input tables and hence, are permutable. The number of distinct binary tree shapes -- denoted as $b(V)$ -- extracted from the join graph $G$ is calculated by the Catalan numbers~\cite{Cormen:ALGORITHMS:mit-2022}. Since each vertex can be any of the input tables, there are $|V|!$ ways to assign the tables to the vertices for each tree shape. Thus, the total size of the search space is:

\begin{equation}
	\begin{split}
		t_b(V) = \frac{(2 \times |V|)!}{(|V| + 1)!}
	\end{split}
	\label{eq:binary-tree-no}
\end{equation}

The number of binary tree shapes growths exponentially~\cite{Cormen:ALGORITHMS:mit-2022} with $V$ including cross-joins -- which is $\Omega\left((4^{|V|}) / |V|^{3/2}\right)$. Cross-join is to join two tables or connected components over a non-existent edge $e \notin E$ in $G$. For example, the join order $(mk \Join k \Join cn \Join mc \Join t)$ exhibits the cross-join between subplan $(mk \Join k)$ and table $cn$. Since a cross-join does not require any condition to match data between two tables, it combines every row of one table with every row of the other table. Despite the possibility that a query plan involving cross-joins can potentially lead to an optimal plan, cross-joins are typically avoided due to their high costs. Thus binary trees that involve cross-joins can be considered a separate subspace. Nonetheless, even this reduced search space still poses a computational challenge due to its vastness.

\textbf{Search space in terms of graph edges.}
In this work, we define the search space over joins -- edges $e \in E$ in $G$ -- which directly excludes cross-joins. The size of the search space is given by the number of spanning trees extracted from the join graph $G$ which is upper bounded by the number of ordered edge arrangements, of size $(|V| - 1)$, selected from $E$:

\begin{equation}
	\begin{split}
		t_s(V, E) & \leq \frac{|E|!}{(|E|-|V|+1)!}
	\end{split}
	\label{eq:span-tree-no}
\end{equation}

The reason for having ordered arrangements instead of only combinations~\cite{Arndt:ARRANGEMENTS:2010} is because the order in which edges are selected matters, described in Section~\ref{subsec:cost-model}, and different edge orders result in different join orders, thus, different query plans. This results in a considerable reduction in the number of binary trees. In the case of a query without cycles -- a join tree with $(|V| - 1)$ edges -- the bound is tight since $|E| = |V| - 1$ and the number of spanning trees is $t_s(V, E) = (|V| - 1)!$ where $0!$ to be equal to $1$. The value of the bound increases with the number of joins, reaching its maximum value for a clique query with $|E| = |V| \times (|V| - 1) / 2$ edges.

The number of spanning trees $t_s(V, E)$ increases at a slower rate than the bound because of the redundancy incurred by cycles. In this case, many arrangements result in invalid spanning trees that cover only a subset of the tables and include cycles. In Figure~\ref{fig:2a_query}, an example of such arrangement is $\{(\textit{mk}-\textit{k}), (\textit{t}-\textit{mk}), (\textit{t}-\textit{mc}), (\textit{mk}-\textit{mc})\}$. The exact value of $t_s(V, E)$ depends heavily on the topology of the join graph. For our example query, $t_s(V, E)$ is $72$ while the bound is $5! = 120$. The rest is the number of invalid query plans -- $48$ invalid spanning trees. Thus, $t_s(V, E)$ is only $60\%$ of the bound. Although this may seem small, as the number of tables increases, so does $t_s(V, E)$, resulting in a large search space that becomes even larger when taking into account the availability of indexes and the types of join algorithms. Thus, given the factorial size of both search spaces, finding an optimal query plan remains a challenging problem~\cite{Ibaraki:NP:tds-1984}.

\begin{figure}[htbp]
	\centering
	\includegraphics[scale=0.85]{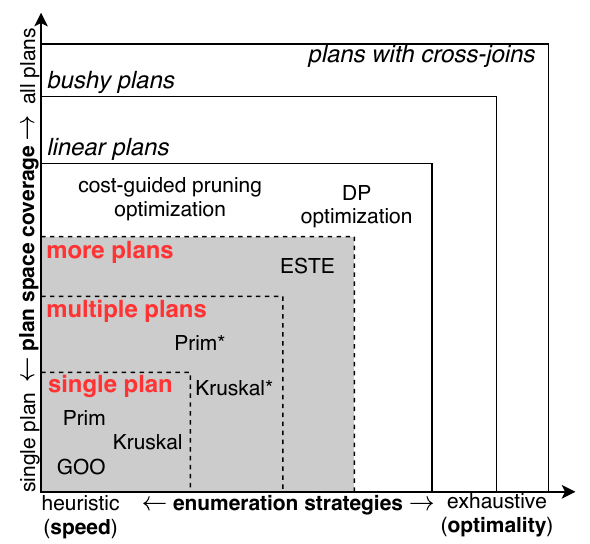}
	\caption{Enumeration strategies and their ability to cover the plan search space. There is no position difference on the y-axis for enumeration methods within the same rectangle.}
	\label{fig:enum_strategies}
\end{figure}

\textbf{Linear search space.}
A standard approach for reducing the size of a given search space is to constrain the shape of query plans~\cite{Moerkotte:DP:sigmod-2008,Moerkotte:TDDP:pvldb-2013}. This means only considering spanning trees of a certain shape. Bushy and linear trees are two principal shapes of query plans. Focusing on specific types of trees helps to streamline the plan search and manage the size of the search space, thereby simplifying the discovery of an optimal query plan. However, this reduction in search space may inadvertently omit optimal query plans. The objective, therefore, is to find a query plan -- optimal plan or near the globally optimal query plan -- within the reduced search space. This reduction balances the computational feasibility of the plan search against the query plan optimality. In order to limit the search space to linear query plans, a single connected component -- a single subtree -- without cycles must be maintained. Thus every vertex has to be recursively considered at each step to ensure that both vertices of an edge are not a member of the current single connected component. We repeat this procedure until we cover all the vertices, making sure that we always maintain a single connected component without cycles. In Figure~\ref{fig:2a_query_plans}, the leftmost and rightmost query plans are trees from the linear search space. The middle query plan has two connected components -- two subtrees, $(cn \Join mc \Join t)$ and $(k \Join mk)$, joined at the root of the tree. The number of the spanning trees generated from our example query is $72$ including $36$ linear and $36$ bushy query plans. Although the number of query plans considered in the linear search space can be smaller, it does not mean the optimal plan cannot be selected. In the case of an optimal plan having a linear shape, reducing the initial search space to the linear search space does not affect the result but can reduce the search time. For example, in Figure~\ref{fig:2a_query_plans}, the leftmost tree is an optimal query plan that has a linear shape.

\subsection{Plan Enumeration}
Within the scope of a given search space, described in Section~\ref{subsec:search-space}, plan enumeration generates and evaluates a vast number of query plans. Each query plan is semantically equivalent but with a different cost. The objective is to find an optimal query plan, which entails the minimum cost. In the context of spanning trees, finding an optimal query plan is equivalent to finding a minimum spanning tree. In this section, we outline common enumeration strategies, while describing the balance between discovering optimal plans and preserving computational efficacy~\cite{Lan:SURVEY:dse-2021,Steinbrunn:HRO:vldb-1997}. Figure~\ref{fig:enum_strategies} illustrates search space coverage by different enumeration strategies. The x-axis presents enumeration strategies from heuristic strategy (faster search time) to exhaustive strategy (preserved plan optimality). The y-axis describes the search space coverage from a single to all query plans. We use the area of rectangles of different sizes to describe the correlation between the two axes. The largest solid rectangle covers all possible query plans including the plans with cross-joins. The second and third largest solid rectangles cover bushy and linear tree shapes, respectively.

\textbf{Exhaustive Enumeration.}
A straightforward and most expensive strategy is to generate every query plan and compute its cost -- covering the largest solid rectangle in Figure~\ref{fig:enum_strategies}. This strategy guarantees the optimality of the selected plan. However, due to the combinatorial explosion of possible query plans, this naive exhaustive enumeration strategy becomes computationally prohibitive for large queries. To expand the feasibility limits of exhaustive strategy for large queries, cross-joins are excluded~\cite{Moerkotte:ATE:vldb-2006,Moerkotte:DP:sigmod-2008} and the search space is limited to linear tree shapes~\cite{Selinger:APSRDMS:sigmod-1979} -- targeting the second and third largest solid rectangles in Figure~\ref{fig:enum_strategies}. Additionally, the computational bottleneck of exhaustive strategy is mitigated through optimizations such as dynamic programming and cost-based pruning~\cite{Selinger:APSRDMS:sigmod-1979,Vance:RBJ:sigmodrec-1996,Moerkotte:ATE:vldb-2006,Moerkotte:DP:sigmod-2008,Dehaan:TOPDOWN:sigmod-2007,Fender:GBTOPDOWN:icde-2011}. Although these optimization techniques reduce the number of query plans evaluated and ensure the discovery of an optimal query plan in a given search space, their efficacy is constrained when applied to large queries.

\textbf{Heuristic Enumeration.}
Exhaustive enumeration strategies can be simplified by heuristic strategies to partially explore a given search space. Thus it further reduces the number of query plans to be generated and evaluated. For example, greedy algorithms~\cite{Swami:LJQ:sigmod-1989,Steinbrunn:HRO:vldb-1997,Fegaras:NEWHEU:dexa-1998,Bruno:POLYNOM:icde-2010,Neumann:QSIMPLE:sigmod-2009,Kossmann:IDP:tods-2000} directly compute a single query plan -- depicted in the smallest dotted rectangle in Figure~\ref{fig:enum_strategies}. Even though the decision at every step is locally optimal, there is no guarantee that the final query plan has the minimum cost among all query plans. This is due to conditioning the available choices at a step on previous decisions. Figure~\ref{fig:2a_query_plans} exhibits this issue by comparing optimal query plan $\mathcal{P}_{opt}$ to query plans $\mathcal{P}_{kru}$ and $\mathcal{P}_{pri}$. Their overall query plan costs are $\mathcal{C}(\mathcal{P}_{\textit{opt}}, Y) = 1.6M$, $\mathcal{C}(\mathcal{P}_{\textit{kru}}, Y) = 2.4M$ and $\mathcal{C}(\mathcal{P}_{\textit{pri}}, Y) = 4.6M$, respectively. The exhaustive enumeration finds the optimal query plan $\mathcal{P}_{opt}$ based on reaching all distinct $14$ subplans, computing all distinct $32$ join costs and comparing all $72$ query plans -- excluding the cross-joins. However, for the query plans $\mathcal{P}_{kru}$ and $\mathcal{P}_{pri}$, 3-way joins $\mathcal{P}_{mk \Join k \Join mc}$, $\mathcal{P}_{mk \Join k \Join t}$, $\mathcal{P}_{mk \Join t \Join mc}$ are never considered. In 4-way joins, $\mathcal{P}_{mk \Join k \Join mc \Join t}$ and $\mathcal{P}_{mk \Join k \Join mc \Join cn}$ are also never reached. The former 4-way join has significantly less cost and intermediate table cardinality which is reached by the exhaustive enumeration. As a result, the greedily selected query plans tend to be suboptimal compared to their exhaustive counterparts. At the same time, considering fewer subplans and building a query plan bottom-up means relying on estimates of smaller join size, which are -- in principle -- more accurate due to the reduced correlation between tables~\cite{Leis:JOB:vldb-2018}. Thus, while exhaustive enumeration requires consistent estimation across all join sizes, the greedy algorithm is more sensitive to simpler estimates. Therefore, the reduction in the search space enumeration can be compensated by consistent estimation of small joins.
\section{SPANNING TREE-BASED ENUMERATION}\label{sec:st_algorithms}
Finding the Minimum Spanning Tree (MST) is to determine a tree from a given graph that spans all the vertices while ensuring that the cumulative cost of its edges is the lowest possible. This problem is a fundamental topic in graph theory and has broad applications in various fields, including query optimization in databases. To find the minimum spanning tree, traditional minimum spanning tree algorithms, such as Prim's and Kruskal's~\cite{Cormen:ALGORITHMS:mit-2022}, operate under the assumption that edge weights are fixed and do not change once defined -- otherwise, optimality is not guaranteed. Essentially, these algorithms maintain an invariant which ensures that the partial MST being constructed at any given time is always a subset of a full MST, upholding the integrity and optimality of the solution throughout the MST construction. It is noteworthy to mention that there can be multiple optimal query plans -- several minimum spanning trees -- within a given search space. This case occurs in graphs that contain cycles or edges with the same edge weights.

In the context of query optimization, we cannot assign static weights to edges in the join graph and then sum up the edge weights in a spanning tree due to the fact mentioned in Section~\ref{subsec:cost-model}. These static weights fail to encapsulate the dynamic effects of prior subplan joins -- multi-table correlation. This precludes the direct application of well-known minimum spanning tree algorithms such as Prim and Kruskal as they fall short in their capacity to efficiently accommodate dynamic changes in edge weights. For example, in Figure~\ref{fig:2a_query_plans}, we observe that the edge $(\textit{mk}-\textit{k})$ appears in all three spanning trees. Its weight is $1.1M$ in $\mathcal{P}_{\textit{opt}}$ and $\mathcal{P}_{\textit{kru}}$, and $0.2K$ in $\mathcal{P}_{\textit{pri}}$. This is due to the different edge subsequences that precede $(\textit{mk}-\textit{k})$. The weight of an edge is derived from the edge subsequence -- or query subplan -- it is appended to. Therefore, given the crucial roles of both the sequence in which edges are arranged and the variable characteristics of edge costs, we refine the problem statement as -- \textit{an optimal query plan is a spanning tree, where the objective is to identify an ``ordered sequence'' of edges spanning all vertices in the join graph while minimizing the total edge weights which ``dynamically change as tables are successively joined''.}

\begin{figure*}[htbp]
	\centering
	\includegraphics[scale=0.75]{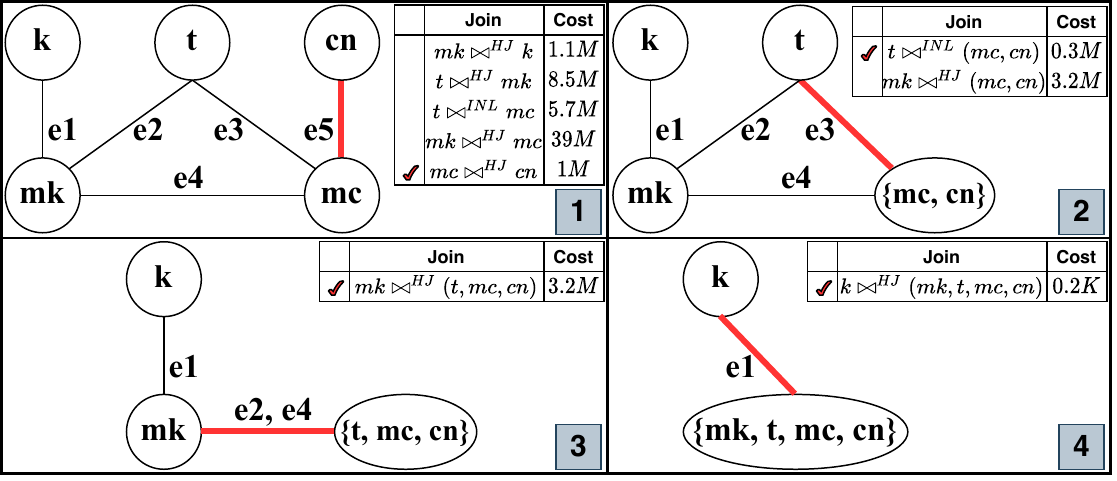}
	\caption{Step-by-step illustration of Prim's plan enumeration algorithm on the join graph of query 2a.}
	\label{fig:2a_prim}
\end{figure*}

\subsection{Spanning Tree-based Plan Generation}
In Figure~\ref{fig:2a_query_plans}, we illustrate three spanning trees $\mathcal{P}_{\textit{opt}}$, $\mathcal{P}_{\textit{kru}}$ and $\mathcal{P}_{\textit{pri}}$. At leaf vertices, we show the cardinalities of base tables. For each internal vertex, we compute join cost $\mathcal{C}$, described in Equation~\ref{eq:cost-model}, using true cardinalities $Y$ of base and intermediate tables. For a join operator selection, we first compute the cost for hash join $HJ$ and index nested-loop join $INL$ if one of the child vertices is a base table with pre-built indexes. A join operator with the minimum cost is selected. Each spanning tree is generated from bottom to top following the ordered edge sequence from left to right: $\mathcal{P}_{\textit{opt}} = \{(\textit{mk}-\textit{k}), (\textit{mk}-\textit{mc}), (\textit{mc}-\textit{cn}), (\textit{t}-\textit{mk})\}$, $\mathcal{P}_{\textit{kru}} = \{(\textit{mc}-\textit{cn}), (\textit{t}-\textit{mc}), (\textit{mk}-\textit{k}), (\textit{t}-\textit{mk})\}$ and $\mathcal{P}_{\textit{pri}} = \{(\textit{mc}-\textit{cn}), (\textit{t}-\textit{mc}), (\textit{mk}-\textit{mc}), (\textit{mk}-\textit{k})\}$. To illustrate the generation and cost computation of a spanning tree, we take $\mathcal{P}_{\textit{opt}}$ as an example. The shape of the $\mathcal{P}_{\textit{opt}}$ is a linear tree. The subplan $\mathcal{P}_{mk \Join k} = \{(\textit{mk}-\textit{k})\}$ is a 2-way join having two children as base tables $k$ and $mk$. The cost of $HJ$ and $INL$ joins are $1.1M$ and $10M$, respectively. Thus, $HJ$ is selected as the join operator type for $\mathcal{P}_{mk \Join k}$. The cardinality of the output table for $\mathcal{P}_{mk \Join k}$ is $42K$. The next subplan $\mathcal{P}_{mk \Join k \Join mc} = \{(\textit{mk}-\textit{k}), (\textit{mk}-\textit{mc})\}$ is a 3-way join having two children as subtree $\mathcal{P}_{mk \Join k}$ and base table $mc$. $INL$ is selected with the cost of $0.3M$ while the cost of $HJ$ is $0.7M$. It is important to note the cost of $\mathcal{P}_{mk \Join k \Join mc}$ takes into account the edge subsequence -- the cardinality of the contained 2-way join subplan $\mathcal{P}_{mk \Join k}$. Thus this edge cost is different than static weight on the edge. The 4-way join $\mathcal{P}_{mk \Join k \Join mc \Join cn} = \{(\textit{mk}-\textit{k}), (\textit{mk}-\textit{mc}), (\textit{mc}-\textit{cn})\}$ has two children as subtree $\mathcal{P}_{mk \Join k \Join mc}$ and base table $cn$. Their $HJ$ and $INL$ join costs are $0.2M$ and $0.3M$, respectively. $HJ$ is selected and the adjacent edge weights are changed accordingly. Lastly, the final 5-way join $\mathcal{P}_{mk \Join k \Join mc \Join cn \Join t} = \{(\textit{mk}-\textit{k}), (\textit{mk}-\textit{mc}), (\textit{mc}-\textit{cn}), (\textit{t}-\textit{mk})\}$ has two children as subtree $\mathcal{P}_{mk \Join k \Join mc \Join cn}$ and base table $t$. $INL$ is selected as final join operator -- costs are $HJ = 0.5M$ and $INL = 16K$, respectively. The remaining edge $(\textit{t}-\textit{mc})$ creates a cycle. Thus join predicate \textit{t.id = mc.movie\_id} can be evaluated as a filter with $(\textit{t}-\textit{mk})$ which is join predicate \textit{t.id = mk.movie\_id}.

\subsection{Exhaustive Spanning Tree-based Enumeration}\label{subsec:exhaustive-st-enum}
Due to dynamic alteration of edge weights resulting from ongoing joins, standard minimum spanning tree algorithms are not guaranteed to yield optimal query plans. Hence, to uncover an optimal query plan, it becomes imperative to enumerate every possible spanning tree over the join graph and select the one with the minimum cost. Rather than employing an exhaustive arrangement-based enumeration strategy with subsequent cost assessments, it is more beneficial to use a recursive algorithm that leverages backtracking. This approach allows to avoid cross-joins and cycles and apply cost-based pruning and dynamic programming optimizations. Every feasible edge has to be recursively considered at any step of the traversal and if the cost of the current subplan surpasses the current minimal cost, it allows for a backtrack. This process aids in circumventing parts of the search space that are not necessary to explore, thus enhancing efficiency by preventing wasteful computations. Employing dynamic programming systematically builds upon subplans to avoid redundant cost computations and adapt to the dynamic weight changes in edge costs. Not only these optimizations may reduce the number of plans considered, but it is guaranteed that the spanning tree with minimum cost is found.

\subsection{Heuristic Spanning Tree-based Enumeration}\label{subsec:heuristic-st-enum}
The intrinsic ability of edge-based enumeration to preclude cross-joins in the search space of the spanning tree presents a substantial advantage. However, even with the optimizations described in Section~\ref{subsec:exhaustive-st-enum}, exhaustive spanning tree-based enumeration can still be computationally prohibitive when dealing with large queries. Hence, a natural step forward is to adapt traditional minimum spanning tree algorithms, such as Prim's and Kruskal's, to accommodate the unique characteristics of a join graph -- dynamic nature of edge weight changes. In addition to their polynomial time complexity, these greedy algorithms provide tangible advantages when dealing with cardinality estimates. Their bottom-up approach in building a query plan emphasizes smaller join size estimates -- large join size estimates are generally far less accurate than smaller joins~\cite{Leis:JOB:vldb-2018}. These aspects make a greedy strategy pragmatic, balancing the need for accuracy and practical effectiveness. We show adapted Prim's and Kruskal's algorithms that take into account the order of an edge sequence and the dynamic change of edge weights. While they may not provide a minimum spanning tree, these algorithms deliver spanning trees with low costs.

\textbf{Prim's Plan Enumeration.}
Prim's algorithm operates within a linear search space of spanning trees, where cross-joins are naturally eliminated. Prim's greedy decision-making nature further reduces the search space by enumerating a single complete spanning tree of a linear shape. At each iteration, Prim's algorithm maintains a single main component and progressively builds a spanning tree -- an ordered sequence of edges. In Figure~\ref{fig:2a_prim}, we demonstrate an adaptation of Prim's algorithm that takes into account changes in edge weights. First, an edge of a 2-way join with the minimum cost is selected from the list of all edges corresponding to 2-way joins -- table in step 1 of Figure~\ref{fig:2a_prim}. The size of the edge list is at most $E$ and is not necessarily sorted as the composition of this list is heavily reliant on previous join selection and their weights are updated at each iteration. Thus, an auxiliary data structure is not necessary. For each join, the optimal join operator is selected as a hash join $HJ$ or indexed nested-loop $INL$ depending on their computed costs as in Equation~\ref{eq:cost-model}. In step 1, edge $e5$, subplan $\mathcal{P}_{mc \Join cn} = \{(\textit{mc}-\textit{cn})\}$, is selected as 2-way $HJ$ with the minimum cost $\mathcal{C}(\mathcal{P}_{mc \Join cn}, Y) = 1M$. The vertices $mc$ and $cn$ form a combined component $\{mc, cn\}$ shown in step 2. The weights of the edges adjacent to vertices $mc$ and $cn$ are updated -- edges $e3$ and $e4$. Next, an edge of a 3-way join with the minimum cost is selected from the list of two edges $e3$ and $e4$ which are adjacent to the main component $\{mc, cn\}$. Edge $e3$, subplan $\mathcal{P}_{mc \Join cn \Join t} = \{(\textit{mc}-\textit{cn}), (\textit{t}-\textit{mc})\}$, is selected as 3-way $INL$ with the minimum cost $\mathcal{C}(\mathcal{P}_{mc \Join cn \Join t}, Y) = 0.3M$. The component $\{mc, cn\}$ and vertex $t$ form a combined component $\{mc, cn, t\}$ shown at step 3. In step 3, the weights of edges $e2$ and $e4$, adjacent to $\{mc, cn, t\}$, are updated. Note that both edges have the same cost of $3.2M$ as a hash join, and either of the edges becomes a cyclic edge. Thus, we randomly select an edge, $e2$ or $e4$, and transform the other edge into a selection predicate as described in Section~\ref{sec:spanning_tree}. After building the subplan $\mathcal{P}_{mc \Join cn \Join t \Join mk} = \{(\textit{mc}-\textit{cn}), (\textit{t}-\textit{mc}), (\textit{mk}-\textit{mc})\}$, the cost of the remaining edge $e1$, adjacent to the main component $\{mc, cn, t, mk\}$, is updated and selected as 5-way $HJ$. The resulting plan $\mathcal{P}_{mc \Join cn \Join t \Join mk \Join k} = \{(\textit{mc}-\textit{cn}), (\textit{t}-\textit{mc}), (\textit{mk}-\textit{mc}), (\textit{mk}-\textit{k})\}$ spans all the vertices in the join graph. In Figure~\ref{fig:2a_query_plans}, the rightmost tree $\mathcal{P}_{pri}$ is the spanning tree generated using Prim's enumeration.

\begin{algorithm}[htbp]
	\begin{multicols}{2}
	\begin{flushleft}
		\textbf{Input}: $\mathcal{S}_{pri} \gets \emptyset$ // \textit{main component} \\
		\tab\tab $\mathcal{S} \gets \{v \mathrel{}\mid\mathrel{} v \in V\}$ // \textit{complement component} \\ 
		\tab\tab $\mathcal{E} \gets \{e : \mathcal{C}(\mathcal{P}_{e}) \mathrel{}\mid\mathrel{} e \in E\}$ // \textit{subplans and costs} \\
		\textbf{Output}: $\mathcal{P}_{pri} \gets []$ // \textit{query plan}
	\end{flushleft}
	\begin{algorithmic}[1]
        \While{$\mathcal{E} \neq \emptyset$}
            \State // \textit{find a join with the minimum cost}
            \State find $e \mathrel{}\mid\mathrel{} \mathcal{C}(\mathcal{P}_{e}) < \mathcal{C}(\mathcal{P}_{e^\ast}), \forall e^\ast \in \mathcal{E}$
            \State $\mathcal{E} \gets \mathcal{E} \mathbin{/} e$
            \State add $e$ to $\mathcal{P}_{pri}$

            \State // \textit{either of $v1_e$ and $v2_e$ is a base table}
            \State $\mathcal{S}_{pri} \gets \mathcal{S}_{pri} \cup \{v1_e, v2_e : |v_e| = 1\}$
            \State $\mathcal{S} \gets \mathcal{S} \mathbin{/} \{v1_e, v2_e : |v_e| = 1\}$

            \For{each $e' \in \mathcal{E}$}
                \If{$v1_{e'} \notin \mathcal{S}_{pri} \land v2_{e'} \notin \mathcal{S}_{pri}$}
                    \State // \textit{remove edges nonadjacent to $\mathcal{S}_{pri}$}
                    \State $\mathcal{E} \gets \mathcal{E} \mathbin{/} e'$
                \ElsIf{$v1_{e'} \in \mathcal{S}_{pri} \land v2_{e'} \in \mathcal{S}_{pri}$}
					\State // \textit{cyclic edge}
                    \State add $e'$ to $\mathcal{P}_{pri}$ as a filter predicate
                \Else ~// \textit{update edges adjacent to $\mathcal{S}_{pri}$ and costs}
                    \State $\mathcal{E} \gets e' \cup \mathcal{S}_{pri} : \mathcal{C}(\mathcal{P}_{e' \cup \mathcal{S}_{pri}})$
                \EndIf
            \EndFor
        \EndWhile
	\end{algorithmic}
	\end{multicols}
	\caption{Prim's Algorithm for Plan Enumeration}
	\label{alg:prim_algorithm}
\end{algorithm}

Prim's plan enumeration is given in Algorithm~\ref{alg:prim_algorithm}. It starts with an empty set $\mathcal{S}_{pri}$ and its complement $\mathcal{S}$ of size $|V|$. 2-way join edges along with their pre-computed costs are given in $\mathcal{E}$. Until $\mathcal{E}$ becomes empty, at each iteration (lines 1-16), an edge $e$ of the current join size, adjacent to $\mathcal{S}_{pri}$, and with the minimum cost is selected (lines 2-5). A min-heap is not necessary since the list of edges must be adjacent to $\mathcal{S}_{pri}$ and their weights are updated at each iteration. In lines 6-8, $\mathcal{S}_{pri}$ and $\mathcal{S}$ are updated with respect to $e$. In lines 9-16, the new costs of edges adjacent to $\mathcal{S}_{pri}$ are updated. Cyclic edges are transformed into filter predicates (lines 13-14). In lines 12-14, the edges that are no longer adjacent to $\mathcal{S}_{pri}$ are dropped. The algorithm executes in polynomial time running in $\mathcal{O}\left(|E|^2\right)$ and allocating $\mathcal{O}\left(|V| + |E|\right)$ memory.

\begin{figure*}[htbp]
	\centering
	\includegraphics[scale=0.75]{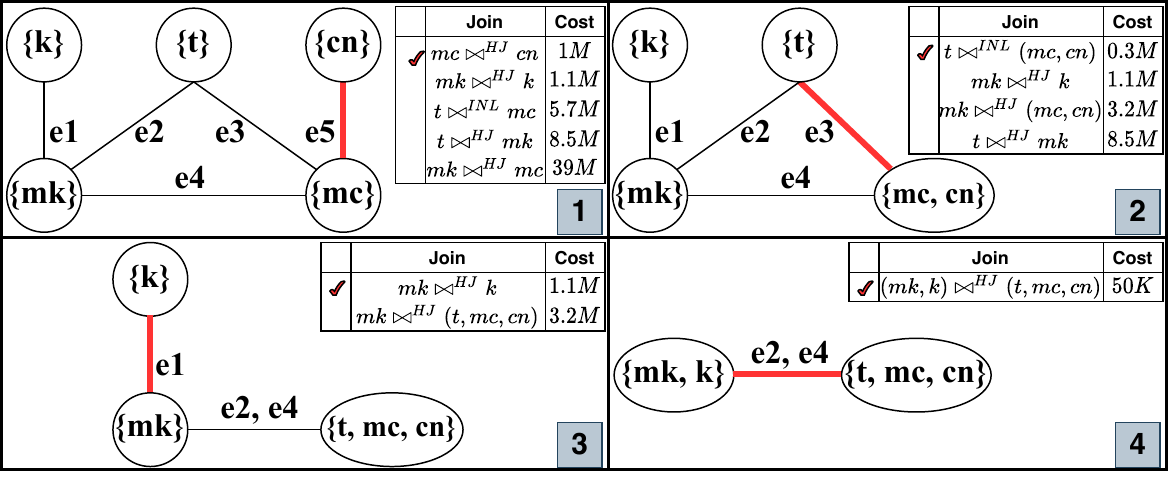}
	\caption{Step-by-step illustration of Kruskal's plan enumeration algorithm on the join graph of query 2a.}
	\label{fig:2a_kruskal}
\end{figure*}

\textbf{Kruskal's Plan Enumeration.}
Kruskal's algorithm is an enhancement over Prim's enumeration by operating within a larger spanning tree search space including bushy and linear tree shapes. Due to its inherently greedy character, Kruskal's enumeration also partially explores the search space and enumerates a single spanning tree. The selected spanning tree has either a bushy or linear tree shape. At each iteration, Kruskal's algorithm combines two separate components of different sizes. Merging two components both larger than one vertex is indicative of a bushy tree. In Figure~\ref{fig:2a_kruskal}, we show a variation of Kruskal's algorithm that takes into account changes in edge weights in selecting an ordered sequence of edges. As in Prim's, first, an edge of a 2-way join with the minimum cost is selected from the list of all edges corresponding to 2-way joins -- table at step 1 of Figure~\ref{fig:2a_kruskal}. Initially, each vertex is a separate component, and for each join, the optimal join operator is selected as either a hash join $HJ$ or indexed nested-loop $INL$ depending on their costs computed as in Equation~\ref{eq:cost-model}. The list of edges, of size at most $E$, is sorted based on their costs. We use a min-heap data structure to maintain the sorted list of edges. The first edge, $e5$ corresponding to subplan $\mathcal{P}_{mc \Join cn} = \{(\textit{mc}-\textit{cn})\}$, is selected as 2-way $HJ$ with the minimum cost $\mathcal{C}(\mathcal{P}_{mc \Join cn}, Y) = 1M$. The vertex components $\{mc\}$ and $\{cn\}$ form a combined component $\{mc, cn\}$ shown at step 2. As in Prim's, the weights of edges $e3$ and $e4$, adjacent to $\{mc, cn\}$, are updated. Unlike Prim's, edges $e1$ and $e2$ are kept with their unchanged costs -- are not adjacent to $\{mc, cn\}$. The selection of these edges leads to bushy trees. In the table at step 2, we list these four edges that can be selected in the next join. Next, an edge of a 3-way or 2-way join with the minimum cost is selected. Edge $e3$, subplan $\mathcal{P}_{mc \Join cn \Join t} = \{(\textit{mc}-\textit{cn}), (\textit{t}-\textit{mc})\}$, is selected as 3-way $INL$ with the minimum cost $\mathcal{C}(\mathcal{P}_{mc \Join cn \Join t}, Y) = 0.3M$. The components $\{mc, cn\}$ and $\{t\}$ form a combined component $\{mc, cn, t\}$ shown at step 3. In step 3, the weights of edges $e2$ and $e4$ adjacent to $\{mc, cn, t\}$ are updated. As in Prim's, both edges of a 4-way join have the same cost of $3.2M$ as a hash join, and either of the edges becomes a cyclic edge. 2-way join $\mathcal{P}_{t \Join mk}$ is dropped from the list since it becomes adjacent to component $\{mc, cn, t\}$ and cannot form a separate subtree. However, edge $e1$ is still kept as it can be selected to form a bushy structure. Since $e1$ has the smallest cost of $1.1M$ compared to edges $e2$ and $e4$ with the cost of $3.2M$, it is selected as a 2-way $HJ$ forming a separate subtree -- subplan $\mathcal{P}_{mk \Join k} = \{(\textit{mk}-\textit{k})\}$. Vertex components $\{mk\}$ and $\{k\}$ form a combined component $\{mk, k\}$ shown at step 4. Lastly, the cost of remaining edges $e2$ and $e4$, adjacent to both components $\{mc, cn, t, mk\}$ and $\{mk, k\}$, are updated. Both edges have the same cost of $50K$ as a hash join, and either of the edges becomes a cyclic edge. We randomly select an edge, $e2$ or $e4$, and transform the other edge into a selection predicate as described in Section~\ref{sec:spanning_tree}. The selected 5-way $HJ$ is a subplan $\mathcal{P}_{mc \Join cn \Join t \Join mk \Join k} = \{(\textit{mc}-\textit{cn}), (\textit{t}-\textit{mc}), (\textit{mk}-\textit{mk}), (\textit{t}-\textit{mk})\}$ which exhibits a bushy shape. The resulting plan spans all the vertices in the join graph. In Figure~\ref{fig:2a_query_plans}, the middle tree $\mathcal{P}_{kru}$ is the spanning tree generated based on Kruskal's enumeration algorithm.

\begin{algorithm}[htbp]
	\begin{multicols}{2}
	\begin{flushleft}
        \textbf{Input}: $\mathcal{S} \gets \{\{v\} \mathrel{}\mid\mathrel{} v \in V\}$ // \textit{$V$ components} \\
            \tab\tab $\mathcal{E} \gets \{\mathcal{C}(\mathcal{P}_{e}) : e \mathrel{}\mid\mathrel{} e \in E\}$ // \textit{min-heap} \\
        \textbf{Output}: $\mathcal{P}_{kru} \gets []$ // \textit{query plan}
	\end{flushleft}
	\begin{algorithmic}[1]
        \While{$\mathcal{E} \neq \emptyset$}
            \State // \textit{extract the first join from the min-heap}
            \State extract $e$ from $\mathcal{E}$

            \State // \textit{find components in $\mathcal{S}$}
            \State $l_e \gets component(v1_e)$
            \State $r_e \gets component(v2_e)$

            \If{$l_e = r_e$} // \textit{cyclic edge}
                \State add $e$ to $\mathcal{P}_{kru}$ as a filter predicate
            \Else // \textit{new component joining $l_e$ and $r_e$}        
                \State $\mathcal{S} \gets \mathcal{S} \cup \{l_e \cup r_e$\}
                \State $\mathcal{S} \gets \mathcal{S} \mathbin{/} \{l_e, r_e$\}
                \State add $e$ to $\mathcal{P}_{kru}$

                \For{each $e' \in \mathcal{E}$}
                    \If{$v1_{e'} \notin \{l_e \cup r_e\} \land v2_{e'} \notin \{l_e \cup r_e\}$}
                        \State // \textit{keep nonadjacent edges}
						\State // \textit{with the same costs}
                    \Else
                        \If{invalid edge}
                            \State // \textit{edges that cannot form a subtree}
							\State // \textit{remove adjacent edges}
                            \State extract $e'$ from $\mathcal{E}$
                        \Else ~// \textit{update adjacent edges and costs}
                            \State $\mathcal{E} \gets \mathcal{C}(\mathcal{P}_{\{e' \cup l_e \cup r_e\}}) : e'$
                        \EndIf                        
                    \EndIf
                \EndFor
            \EndIf
        \EndWhile
	\end{algorithmic}
	\end{multicols}
	\caption{Kruskal's Algorithm for Plan Enumeration}
	\label{alg:kruskal_algorithm}
\end{algorithm}

Kruskal's plan enumeration is given in Algorithm~\ref{alg:kruskal_algorithm}. It starts with a set $\mathcal{S}$ containing $|V|$ separate components. The list of 2-way join edges as values along with their pre-computed costs as keys stored in a min-heap data structure $\mathcal{E}$. Until the set $\mathcal{E}$ becomes an empty set, at each iteration (lines 1-24), the edge $e$ which has the minimum cost is selected -- the first element extracted from the min-heap (lines 2-3). In Kruskal's algorithm, it is efficient to keep a sorted list of edges since some of the joins of lower size are kept which can be selected to build a separate subtree. Hence, we maintain the min-heap data structure to keep edges sorted by cost. In lines 5-7, the components to which the two vertices $v1$ and $v2$ of the current edge $e$ belong are found. Every edge extracted from the min-heap either forms a cycle (lines 9-10) or merges two components (lines 11-24). In the case of cyclic edge, it transforms the edge to a filter predicate. Otherwise, the two components are merged (lines 12-14). In lines 16-24, the min-heap is updated with respect to the selected edge $e$. For demonstration purposes, in lines 17-18, we show the case where the edges that potentially can create a separate subtree are kept with the same costs. Otherwise, an edge that no longer can form a separate subtree is dropped (lines 20-22). In the case of adjacent and valid join, the costs of the corresponding joins are updated (lines 23-24). The algorithm executes in $\mathcal{O}\left(|E|^2 \times log(|E|)\right)$ and allocating $\mathcal{O}\left(|V| + |E|\right)$ memory.

\subsection{Ensemble Spanning Tree-based Enumeration}
\begin{figure}[htbp]
	\centering
	\includegraphics[scale=0.75]{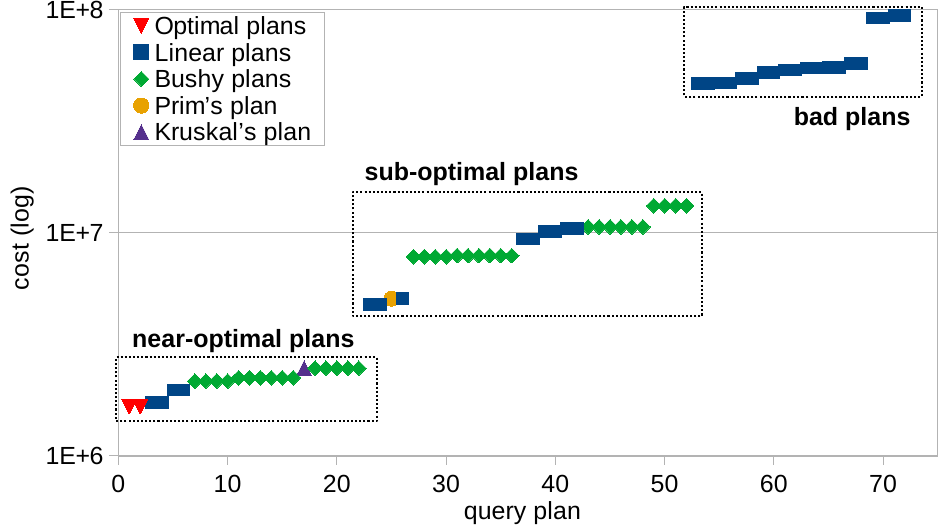}
	\caption{Query plans of query 2a sorted by cost and colored by tree shape. Query plans selected by Prim's and Kruskal's algorithms are shown in orange circles and pink triangles pointed up, respectively.}
	\label{fig:2a_plans_cluster}
\end{figure}

A significant drawback of greedy algorithms lies in their partial enumeration of a given search space -- a single complete plan is enumerated. Moreover, this query plan may get stuck in local optima. This characteristic can be observed in Figure~\ref{fig:2a_plans_cluster}. We plot all spanning trees, excluding cross-joins, generated from the join graph of query 2a and sort them by their cost. Two optimal plans are marked in red triangles located in the bottom left corner. Linear and bushy spanning trees are indicated as blue square and green diamond, respectively. The lowest rectangle shows near-optimal plans which have cost differences of at most $1.31$. The worst query plans are shown inside the upper rectangle -- cost differences of at least $27.9$. Each of these clusters of spanning trees is characterized by their highest join costs, signifying their confinement to local optima. The plans generated by Prim's and Kruskal's algorithms are situated in their respective clusters of spanning trees. From the figure, one can see that limiting the enumeration algorithm to a single full query plan is not effective, as it bears the risk of getting trapped in local optima. Hence, given the limitations of enumerating a single plan and the prohibitive computational expense of exhaustive enumeration, a reasonable approach is to enumerate more than one query plan. This intermediate strategy can help balance the search for optimality with computational efficiency. Enumerating more than one query plan increases the robustness of the search and uncovers more parts of the search space.

A direct method to facilitate the exploration of more query plans is to employ randomized approaches\cite{Steinbrunn:HRO:vldb-1997,Ioannidis:RANDOM:sigmod-1990,Swami:LJQ:sigmod-1989,postgres,Swami:IIO:sigmod-1988,Ioannidis:ANNEAL:sigmod-1987}. In these approaches, the enumeration could randomly begin from a particular edge in the join graph. This kind of stochastic approach could increase the diversity of the examined query plans and help escape local optima. However, rather than starting from random edges, an approach that systematically initiates the enumeration from every edge in the join graph can be a better alternative to preserve the interpretability of the search and ensure more comprehensive coverage of the search space. Unlike prior work~\cite{Krishnamurthy:KBZ:vldb-1986,Ibaraki:IK:tods-1984,Fegaras:NEWHEU:dexa-1998,Lee:MVP:tkde-2001,Moerkotte:BQO-book:2023}, we take one step further by incorporating of a series of various of spanning tree algorithms to further diversify the set of query plans generated. We propose Ensemble Spanning Tree Enumeration (ESTE). This strategy capitalizes on the polynomial-time complexity of spanning tree algorithms while enhancing the robustness of the plan enumeration. Moreover, spanning tree algorithms operate over a simple query representation -- a join graph as opposed to plan graph~\cite{Negi:FLOWLOSS:pvldb-2021,Haffner:HEU:sigmod-2023}. As described in Section~\ref{subsec:heuristic-st-enum}, the memory to maintain the list of edges at each step is at most $E$. Initiating Prim's and Kruskal's enumerations from each edge results in $\mathcal{O}\left(|E|^3\right)$ and $\mathcal{O}\left(|E|^3 \times log(|E|)\right)$ runtime complexity, respectively. Thus ESTE creates an ensemble of spanning trees with low costs broadening the search coverage without escalating the computational cost.

\begin{figure}[htbp]
	\centering
	\includegraphics[scale=0.75]{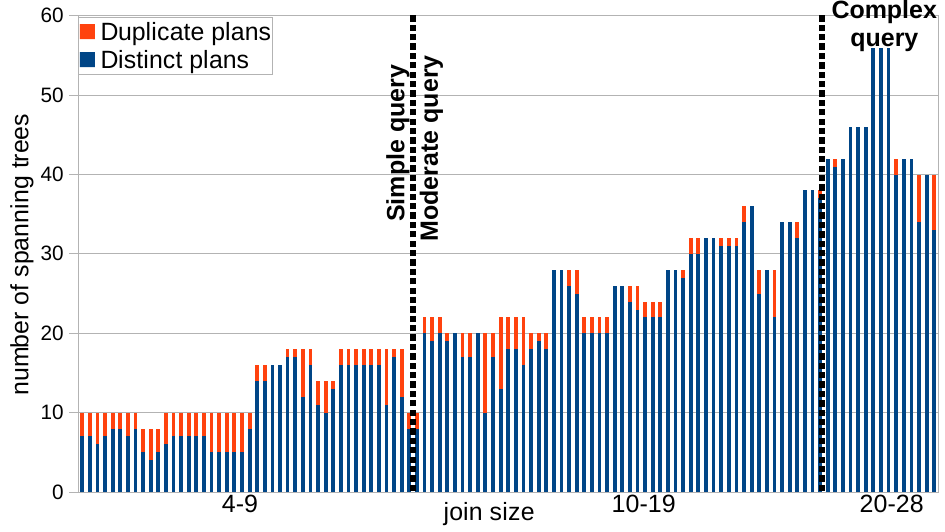}
	\caption{Distinct physical query plans enumerated by ESTE.}
	\label{fig:distinct_plans}
\end{figure}

Based on our example \textit{query 2a}, space coverage by Prim's and Kruskal's single plan enumerations combined reach $10$ or $71.4\%$ out of $14$ distinct subplans, computed $9$ or $28.1\%$ out of $32$ distinct join costs, and $2$ or $2.8\%$ different plans enumerated out of $72$ possible plans excluding the cross-joins. In ESTE, $14$ or $100\%$ subplans, $26$ or $81.2\%$ join costs, and $8$ or $11.1\%$ different plans are enumerated. By making minor trade-offs in optimization time, ESTE explores larger portions of the search space. In Figure~\ref{fig:distinct_plans}, we show distinct physical query plans enumerated by ESTE. The results show a decreasing trend in the number of duplicate physical plans as the complexity of queries increases in number of joins. In Section~\ref{sec:experiments}, we demonstrate that this approach does indeed yield benefits in terms of discovering more efficient plans thus reducing query execution time. The flexibility of ESTE allows for the incorporation of additional spanning tree algorithms in response to changes in data and workload, thus maintaining robustness and even enhancing the optimizer performance. We believe that ESTE offers a more cost-effective way to sustain optimizer robustness, favoring the integration of new algorithms over a complete change and redevelopment of the optimizer for different plan enumeration strategies.
\section{EMPIRICAL EVALUATION}\label{sec:experiments}
Throughout the paper, we provided in-depth micro analysis over the example query \textit{2a}. In this section, we perform a macro analysis over Join Order Benchmark~\cite{Leis:JOB:vldb-2018,Leis:QOREALLY:pvldb-2015} offering a broader performance landscape. We set ESTE against GOO~\cite{Fegaras:NEWHEU:dexa-1998}, one of the most outstanding, if not the best, heuristic enumeration algorithms. Our decision is based on previous evaluations which have consistently demonstrated GOO's superior performance across various configurations, graph topologies, and benchmarks among other proposed heuristic enumeration algorithms~\cite{Leis:JOB:vldb-2018, Leis:QOREALLY:pvldb-2015, Neumann:AOVLJQ:sigmod-2018, Haffner:HEU:sigmod-2023}. Employing GOO as a comparative baseline is expected to provide a solid starting point for our performance assessment. In addition, we evaluate ESTE over different join graph topologies such as chain, cycle, star and clique~\cite{Haffner:HEU:sigmod-2023}. Our evaluation addresses the following questions:

\begin{itemize}[leftmargin=*,noitemsep,nolistsep]
	\item Assess the performance of ESTE -- how it fares against exhaustive enumeration, and how it stands up to cutting-edge heuristic techniques. We evaluate whether ESTE enhances the performance by devoting additional optimization time to exploring a broader portion of the search space.
	\item Separately assess the performance of ESTE over different join graph topologies of large queries.
	\item Assess the performance and response of ESTE when cardinality estimates are used. This helps us understand how ESTE adapts to and performs in the presence of cardinality estimation errors.
\end{itemize}

\subsection{Experimental Setup}\label{subsec:experiments:setup}
The implementation of the current work together with all the experimental artifacts are available online~\cite{este-github}.

\textbf{Dataset \& query workload.}
We perform the experiments on the IMDB dataset~\cite{Boncz:imdb-data}, which has been used extensively to evaluate query optimizers~\cite{Leis:QOREALLY:pvldb-2015,Leis:JOB:vldb-2018,Neumann:AOVLJQ:sigmod-2018,Cai:PCETUB:sigmod-2019,Perron:Worried:icde-2019,Haffner:HEU:sigmod-2023}. JOB benchmark~\cite{JOB-github} defines 113 queries grouped into 33 families. These queries vary significantly in their complexity. The simplest queries have 4 join predicates with the largest join size of 4, while the most complex queries have 28 join predicates with the largest join size of 17. The workload encompasses a diverse range of graph topologies, including chain, star, cycle, and clique. This variability manifests itself in execution times that are highly different. To compensate for this, we split the queries into three complexity groups and examine each group separately. These complexity groups are based on the number of join predicates: simple queries with 4-9 joins, moderate queries with 10-19 joins, and complex queries with 20-28 joins, respectively. We additionally generate different graph topologies including chain, cycle, star, and clique~\cite{mutable-github,Haffner:HEU:sigmod-2023}. The number of relations in these queries is matched as in ~\cite{Haffner:HEU:sigmod-2023}.

\begin{figure*}[htbp]
	\begin{subfigure}[t]{\textwidth}
		\centering
		\includegraphics[scale=0.75]{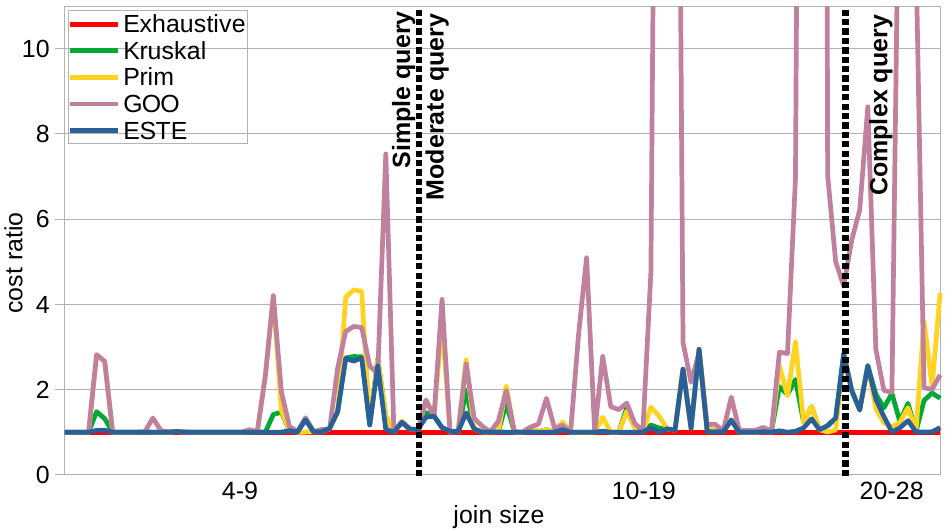}
	  	\caption{True cardinalities}
	  	\label{fig:job_plan_true_cost}
	\end{subfigure}
	\hfill
	\begin{subfigure}[t]{\textwidth}
		\centering
	  	\includegraphics[scale=0.75]{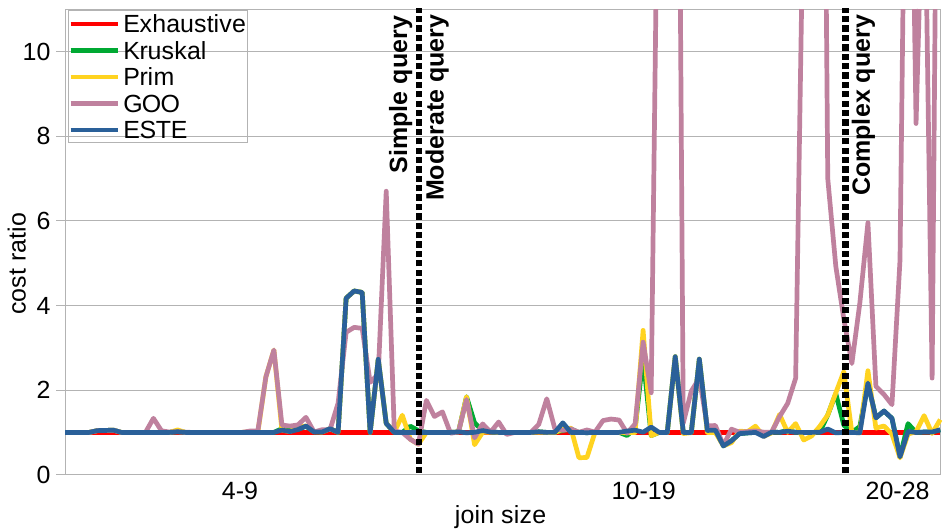}
	  	\caption{Estimated cardinalities}
	  	\label{fig:job_plan_estimated_cost}
  	\end{subfigure}
  	\caption{Cost ratio of selected and optimal query plans in JOB. The costs are computed using true cardinalities.}
  	\label{fig:job_plan_cost}
\end{figure*}

\textbf{Database system \& hardware.}
For cardinality estimations, we use a well-known database system PostgreSQL (version 15.1)~\cite{postgres}. PostgreSQL uses a large variety of statistics on base tables. Any join estimate is computed by combining these statistics into simple arithmetic formulas that make general assumptions on uniformity, inclusion, and independence. We run subqueries of JOB queries in PostgreSQL to collect estimated and true cardinalities. We generate true cardinalities for the join graph topologies as in ~\cite{Haffner:HEU:sigmod-2023}. We use an optimized docker image publicly available for PostgreSQL. We run PostgreSQL with default configurations, and set operator and buffer size to 16GB and 64GB, respectively. For measuring execution time, we utilize \textit{pg\_hint} to force cost function, defined Equation~\ref{eq:cost-model}, which fixes join order as well as physical operators. To compare ESTE against other enumeration algorithms over different graph topologies, we use the cost function as in ~\cite{Haffner:HEU:sigmod-2023}. All the experiments run on an Ubuntu 22.04 LTS machine with 56 CPU cores (Intel Xeon E5-2660), 256GB RAM, and an NVIDIA Tesla K80 GPU.

\begin{figure*}[htbp]
	\begin{subfigure}[t]{\textwidth}
		\centering
		\includegraphics[scale=0.75]{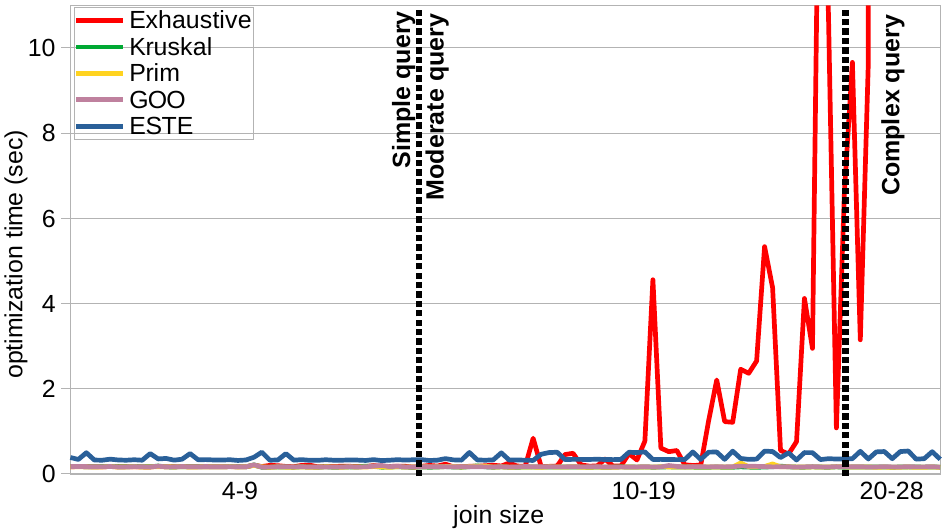}
	  	\caption{True cardinalities}
	  	\label{fig:job_plan_true_opt_time}
	\end{subfigure}
	\hfill
	\begin{subfigure}[t]{\textwidth}
		\centering
	  	\includegraphics[scale=0.75]{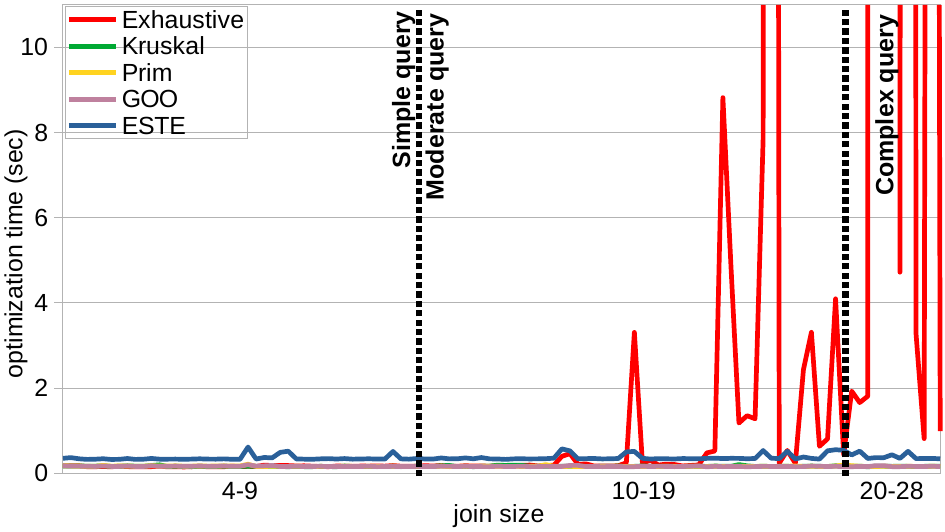}
	  	\caption{Estimated cardinalities}
	  	\label{fig:job_plan_estimated_opt_time}
  	\end{subfigure}
  	\caption{Optimization time (sec) of enumeration algorithms for all JOB query plans.}
  	\label{fig:job_plan_opt_time}
\end{figure*}

\subsection{Evaluation on JOB}
\textbf{Query plan costs.}
In Figures~\ref{fig:job_plan_true_cost} and ~\ref{fig:job_plan_estimated_cost}, we compare query plan costs selected by four heuristic enumeration algorithms using true cardinalities -- Kruskal's, Prim's, GOO, and ESTE. The y-axis shows the cost ratio of heuristic and optimal plans selected by the exhaustive enumeration. The x-axis shows 113 JOB queries grouped by their join complexity as described in Section~\ref{subsec:experiments:setup} -- dotted vertical lines. The enumeration algorithms are illustrated as colored solid lines: exhaustive as red, Kruskal's as green, Prim's as yellow, GOO as light purple, and ESTE as blue. In Figure~\ref{fig:job_plan_true_cost}, the costs are computed using true cardinalities. From the figure, we observe GOO selects relatively the worst query plans. Prim's is the worst among spanning tree-based enumeration algorithms. Prim's limited search in linear search space and constraint to maintain a single connected component forces to explore different areas of the search space. Thus, in certain instances, Prim's chooses superior plans compared to Kruskal's and contributes to enhancing the performance of ESTE. Regardless, ESTE enhances the performance by using both Prim's and Kruskal's and initiating them at each edge. As anticipated, all enumeration algorithms begin to experience a drop in efficiency as the join size increases. In Figure~\ref{fig:job_plan_estimated_cost}, we compare query plans selected using PostgreSQL estimated cardinalities and their costs are computed using true cardinalities. In the figure, we observe similar trends except in the cases when the cost ratios are below 1 which indicate plans that are better than the estimated optimal plans selected by the exhaustive enumeration. This happens when enumeration algorithms are misguided by estimation errors -- underestimations of actual suboptimal plans and overestimations in actual optimal plans. This behavior also can be seen comparing ESTE against Prim's and Kruskal's query plans -- blue line above the yellow and green lines. Overall, ESTE enhances the performance by demonstrating more consistency in plan quality thus decreasing the need for exhaustive enumeration in both scenarios.

\begin{figure*}[htbp]
	\begin{subfigure}[t]{\textwidth}
		\centering
		\includegraphics[scale=0.75]{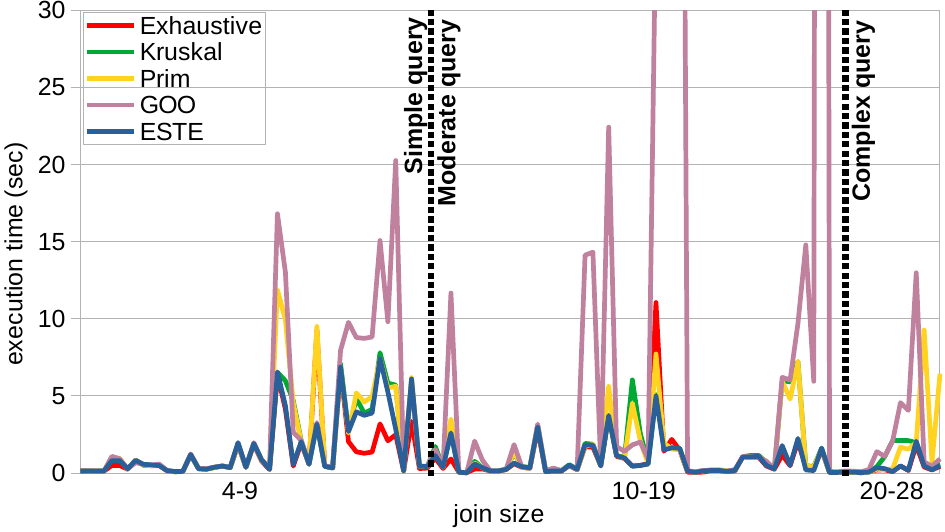}
		\caption{True cardinalities}
	  	\label{fig:job_plan_true_exec_time}
	\end{subfigure}
	\hfill
	\begin{subfigure}[t]{\textwidth}
		\centering
	  	\includegraphics[scale=0.75]{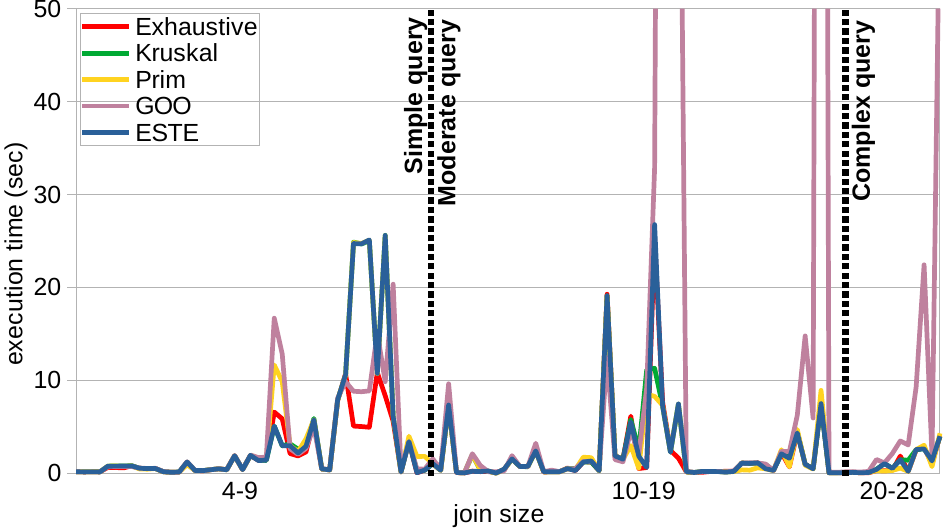}
	  	\caption{Estimated cardinalities}
	  	\label{fig:job_plan_estimated_exec_time}
  	\end{subfigure}
  	\caption{Execution time (sec) of query plans selected by enumeration algorithms.}
  	\label{fig:job_plan_exec_time}
\end{figure*}

\textbf{Optimization time.}
In Figures~\ref{fig:job_plan_true_opt_time} and ~\ref{fig:job_plan_estimated_opt_time}, we evaluate the optimization times of all five enumeration algorithms. The optimization time of the exhaustive enumeration becomes significantly high as the query complexity increases. Because of this, we observe the limit of exhaustive enumeration in large queries -- we were not able to get query plans for family 29 in JOB. ESTE pays minimal extra time in optimization for more consistent and reliable behavior -- maximum $0.54$ sec. Expectedly, Prim's and Kruskal's exhibit very close optimization time to GOO -- all three show optimization time around $0.2$ sec. All spanning tree-based, including ESTE, and GOO enumerations maintain a stable optimization time across different query complexities. In Figure~\ref{fig:job_plan_estimated_opt_time}, we observe a similar trend in the case of estimated cardinalities. The behavior of all algorithms remains similar to true cardinalities except for exhaustive enumeration. As previously mentioned, estimation errors can cause cardinalities to reach saturation resulting in similar plan costs, thus the benefit of the cost-based pruning technique within the exhaustive search diminishes. Consequently, the algorithm is forced to explore a significantly larger part of the search space, which results in a significant increase in optimization time.

\textbf{Execution time.}
Figures~\ref{fig:job_plan_true_exec_time} and ~\ref{fig:job_plan_estimated_exec_time} present the execution times for all five enumeration algorithms. In Figure~\ref{fig:job_plan_true_exec_time}, we observe that GOO's highly suboptimal plans indeed result in longer execution time. Expectedly, the optimal plans selected by the exhaustive enumeration exhibit the best performance except for a small number of exceptions. This is due to the gap between cost function and execution time. Upon examining these cases, we observe that index nested-loop join operators are favored for large tables because of their available indexes. This, in turn, generates large intermediate tables leading to large latencies in the later stages of query execution. In contrast, Prim's and Kurskal's plans perform noticeably better compared to GOO's plans. On the other hand, ESTE brings even more consistency and better performance sacrificing minimal extra time in optimization. Figure~\ref{fig:job_plan_estimated_exec_time} presents execution times for query plans selected using estimated cardinalities. We first observe all enumeration algorithms demonstrate a decline in performance compared to the plans selected using true cardinalities -- larger scale on the y-axis. Just as with true cardinalities, GOO's inaccurate decisions lead to high execution time in large queries. All three spanning tree-based plans show execution times similar to the exhaustive enumeration regardless of query complexity. In several cases, ESTE chooses different plans than Prim's and Kruskal's due to decisions to choose from multiple join options with the same costs. Thus different execution times. To summarize, the results highlight the potential benefits of leveraging a combination of fast-spanning-tree-based heuristics to find the balance between plan quality and computational constraints, especially in complex queries.

\begin{table*}[htbp]
	\centering
	\resizebox{\textwidth}{!}{
	\begin{tabular}{|c|*{15}{c|}} \hline
		\multirowcell{3}{Query \\ Complexity \\ } & \multicolumn{15}{c|}{Enumeration Strategy} \\ \cline{2-16}
		& \multicolumn{3}{c|}{\textbf{Exhaustive}} & \multicolumn{3}{c|}{\textbf{Kruskal}} & \multicolumn{3}{c|}{\textbf{Prim}} & \multicolumn{3}{c|}{\textbf{GOO}} & \multicolumn{3}{c|}{\textbf{ESTE}} \\ \cline{2-16}
		& \multirowcell{2}{cost \\ ratio} & \multirowcell{2}{opt \\ (ms)} & \multirowcell{2}{exec \\ (sec)} & \multirowcell{2}{cost \\ ratio} & \multirowcell{2}{opt \\ (ms)} & \multirowcell{2}{exec \\ (sec)} & \multirowcell{2}{cost \\ ratio} & \multirowcell{2}{opt \\ (ms)} & \multirowcell{2}{exec \\ (sec)} & \multirowcell{2}{cost \\ ratio} & \multirowcell{2}{opt \\ (ms)} & \multirowcell{2}{exec \\ (sec)} & \multirowcell{2}{cost \\ ratio} & \multirowcell{2}{opt \\ (ms)} & \multirowcell{2}{exec \\ (sec)} \\
		& & & & & & & & & & & & & & & \\ \hline
		\textbf{Simple} & 1 & 7.91 & 61 & 1.44 & 7.58 & 86 & 2.09 & 7.47 & 103 & 2.33 & 7.65 & 150 & 1.36 & 15.75 & 75 \\ \hline
		\textbf{Moderate} & 1 & 83.19 & 46 & 1.16 & 8.63 & 70 & 1.45 & 8.89 & 71 & 6.86 & 8.94 & 848 & 1.05 & 20.83 & 44 \\ \hline
		\textbf{Complex} & 1 & 4,171 & 4 & 1.62 & 1.94 & 11 & 2.48 & 1.92 & 22 & 5.66 & 2.0 & 29 & 1.09 & 5.28 & 5 \\ \hline
		TOTAL & 1 & 4,262 & 111 & 1.34 & 18.15 & 167 & 1.86 & 18.28 & 196 & 4.1 & 18.59 & 1,026 & 1.24 & 41.85 & 124 \\ \hline
	\end{tabular}
	}
    \caption{Overall cost ratio of true query plan costs, optimization (ms) and execution (ms) times on JOB.}
	\label{table:true_workload_cost}
\end{table*}

\begin{table*}[htbp]
	\centering
	\resizebox{\textwidth}{!}{
	\begin{tabular}{|c|*{15}{c|}} \hline
		\multirowcell{3}{Query \\ Complexity \\ } & \multicolumn{15}{c|}{Enumeration Strategy} \\ \cline{2-16}
		& \multicolumn{3}{c|}{\textbf{Exhaustive}} & \multicolumn{3}{c|}{\textbf{Kruskal}} & \multicolumn{3}{c|}{\textbf{Prim}} & \multicolumn{3}{c|}{\textbf{GOO}} & \multicolumn{3}{c|}{\textbf{ESTE}} \\ \cline{2-16}
		& \multirowcell{2}{cost \\ ratio} & \multirowcell{2}{opt \\ (ms)} & \multirowcell{2}{exec \\ (sec)} & \multirowcell{2}{cost \\ ratio} & \multirowcell{2}{opt \\ (ms)} & \multirowcell{2}{exec \\ (sec)} & \multirowcell{2}{cost \\ ratio} & \multirowcell{2}{opt \\ (ms)} & \multirowcell{2}{exec \\ (sec)} & \multirowcell{2}{cost \\ ratio} & \multirowcell{2}{opt \\ (ms)} & \multirowcell{2}{exec \\ (sec)} & \multirowcell{2}{cost \\ ratio} & \multirowcell{2}{opt \\ (ms)} & \multirowcell{2}{exec \\ (sec)} \\
		& & & & & & & & & & & & & & & \\ \hline
		\textbf{Simple} & 1 & 7.41 & 102 & 1.38 & 7.38 & 178 & 1.72 & 7.55 & 195 & 1.91 & 7.45 & 152 & 1.38 & 15.99 & 177 \\ \hline
		\textbf{Moderate} & 1 & 185 & 107 & 1.05 & 8.88 & 114 & 1.04 & 8.71 & 102 & 4.25 & 8.55 & 805 & 1.05 & 19.95 & 115 \\ \hline
		\textbf{Complex} & 1 & 24,327 & 13 & 1.02 & 1.97 & 15 & 1.16 & 1.96 & 12 & 14.06 & 1.95 & 110 & 1.02 & 4.72 & 14 \\ \hline
		TOTAL & 1 & 24,520 & 222 & 1.24 & 18.24 & 307 & 1.44 & 18.21 & 309 & 3.21 & 17.95 & 1,067 & 1.24 & 40.65 & 306 \\ \hline
	\end{tabular}
	}
    \caption{Overall cost ratio of estimated query plan costs, optimization (ms) and execution (ms) times on JOB.}
	\label{table:est_workload_cost}
\end{table*}

\textbf{Overall Evaluation}\label{sec:experiments:overall}
Tables~\ref{table:true_workload_cost} and ~\ref{table:est_workload_cost} represent the overall performance of all five enumeration algorithms. The queries, grouped by complexity as described in Section~\ref{subsec:experiments:setup}, are shown in the rows while the performance of each enumeration algorithm is represented in columns. The last row provides the total workload performance for each enumeration algorithm, including its overall cost ratio, and optimization and execution times. In Table~\ref{table:true_workload_cost}, based on true cardinalities, GOO shows relatively the worst overall results across all levels of query complexity. While Prim's and Kruskal's are significantly better at finding more efficient query plans, ESTE enhances this performance even more by probing a larger portion of the search space. While the optimization time exhaustive is significantly high, ESTE spends at most $2.5X$ extra optimization time than Prims's, Kruskal's and GOO. As shown in Figures~\ref{fig:job_plan_cost} and~\ref{fig:job_plan_exec_time}, this consistent extra time investment trades off for consistently better plans, which notably pays off during the query execution. Overall, we observe the execution time of ESTE is superior to Prim's and Kruskal's, and considerably better than GOO's. Interestingly, across all query complexities, the overall execution time of ESTE is noticeably close to the execution time of optimal plans obtained by the exhaustive enumeration.

In Table~\ref{table:est_workload_cost}, we compare the overall workload performances of the enumeration algorithms when operated using PostgreSQL estimated cardinalities. As depicted in Figures~\ref{fig:job_plan_cost} and~\ref{fig:job_plan_exec_time}, the effects of cardinality estimation errors on plan costs and execution time are detrimental. In all enumeration algorithms, including the exhaustive, we observe a $2-2.5X$ times increase in runtime, while GOO's execution time remains high, similar to the results with true cardinalities shown in Table~\ref{table:true_workload_cost}. Although cost ratios seem to decrease compared to true cardinalities, meaning more similarity to the plans obtained by the exhaustive enumeration, their absolute values significantly increased, leading to longer execution times. In optimization time, we observe a prohibitively large increase in the case of exhaustive enumeration due to the closeness of query plan costs affected by estimation errors -- thus the cost-based pruning technique becomes less effective, forcing the enumeration algorithm to explore a much larger plan space. However, similar to true cardinalities, the optimization time of spanning tree-based and GOO algorithms exhibit consistent behavior across all query complexities. To summarize, spanning tree-based enumeration algorithms offer a superior strategy for finding efficient plans, outperforming GOO, one of the existing efficient heuristic enumeration algorithms. By spending minimal extra optimization time, ESTE further noticeably improves the performance over Prim's and Kruskal's enumeration algorithms. This indicates that despite the challenges brought by cardinality estimation errors, it is still possible to achieve comparatively efficient query plans.

\begin{figure*}[htbp]
	\centering
	\begin{subfigure}[t]{0.51\textwidth}
		\includegraphics[width=\textwidth]{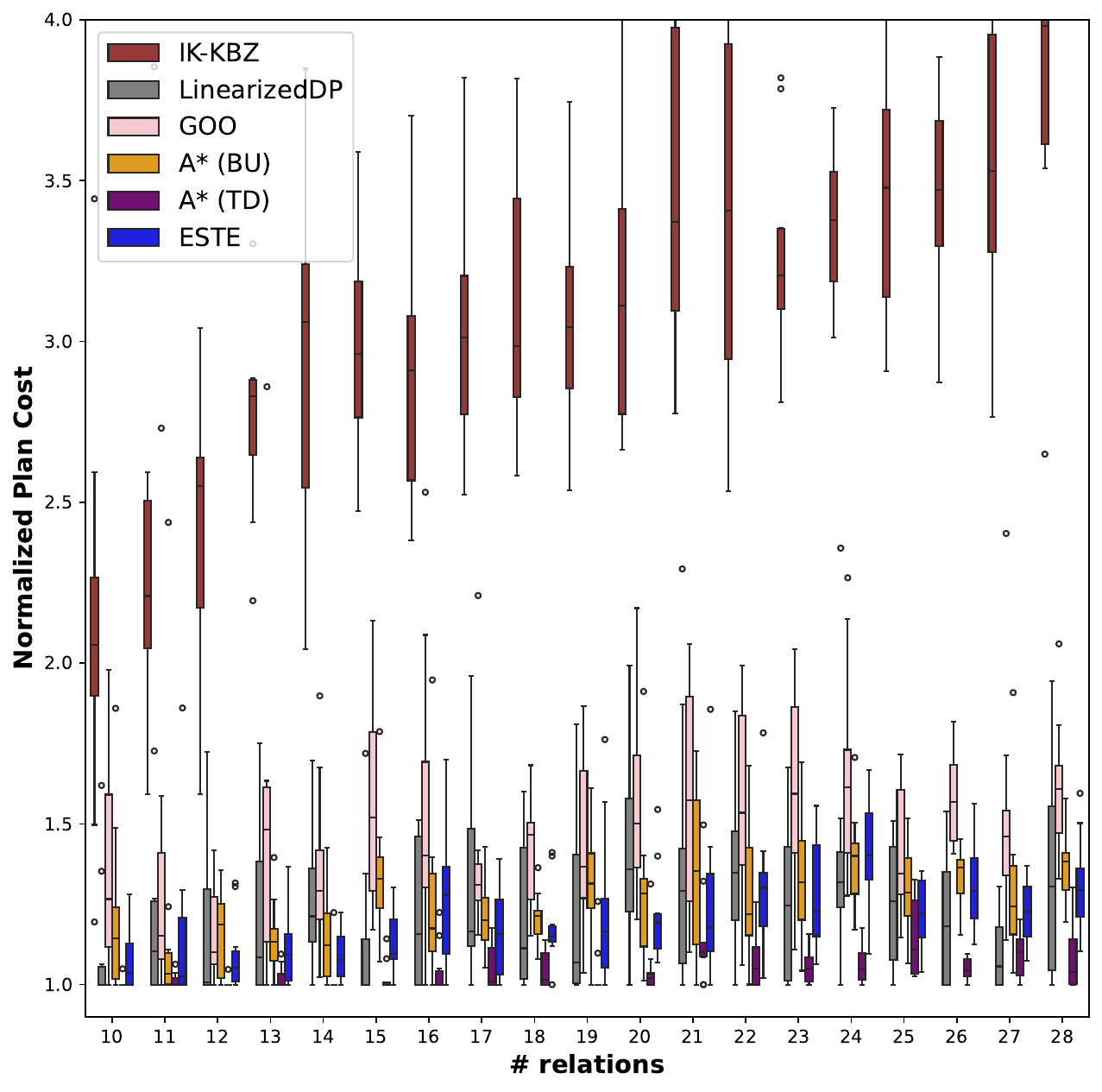}
		\caption{Chain}
	  	\label{fig:top_cout_chain}
	\end{subfigure}
	\begin{subfigure}[t]{0.48\textwidth}
	  	\includegraphics[width=\textwidth]{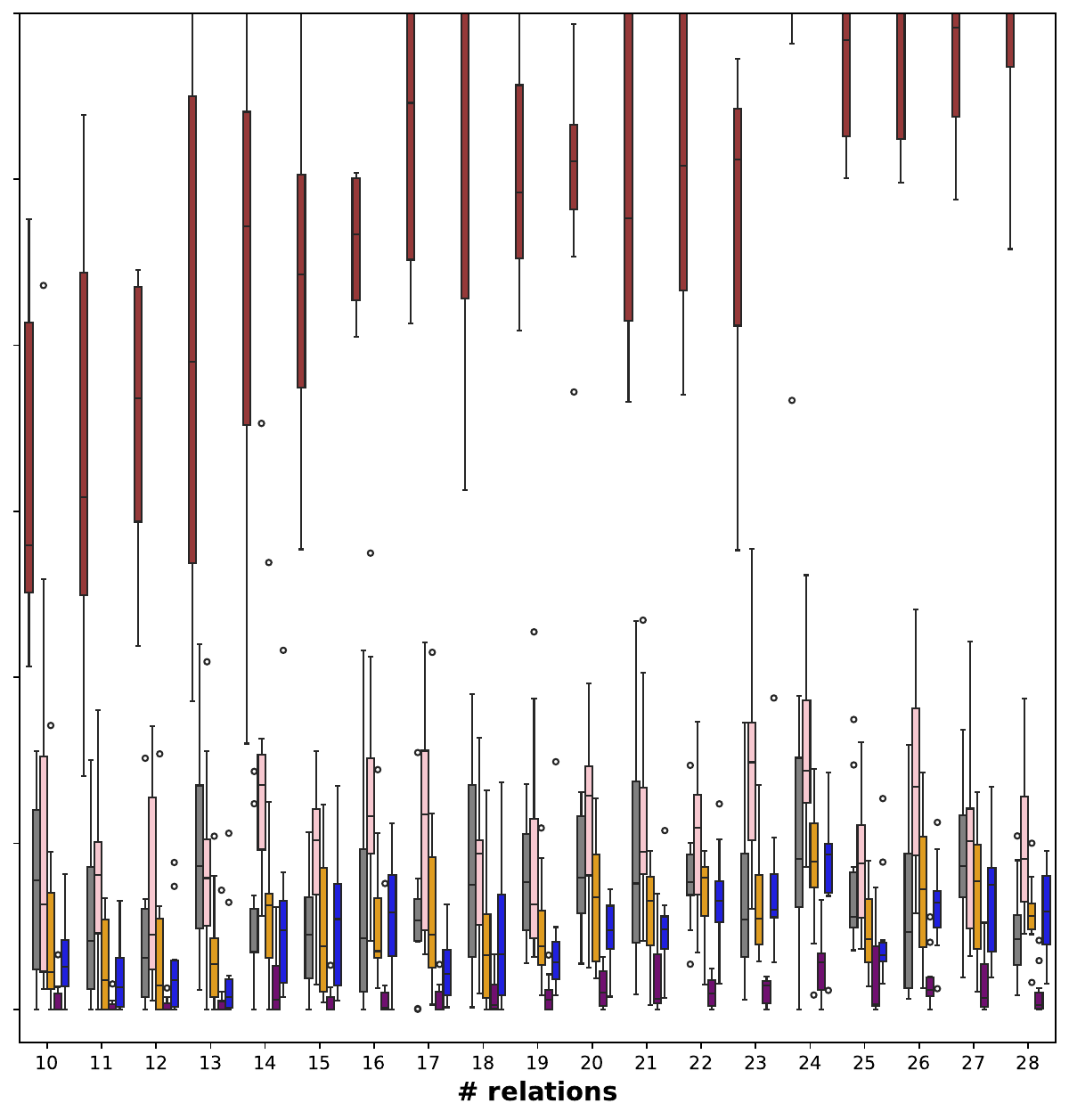}
	  	\caption{Cycle}
	  	\label{fig:top_cout_cycle}
  	\end{subfigure}
	\hfill
	\begin{subfigure}[t]{0.51\textwidth}
		\includegraphics[width=\textwidth]{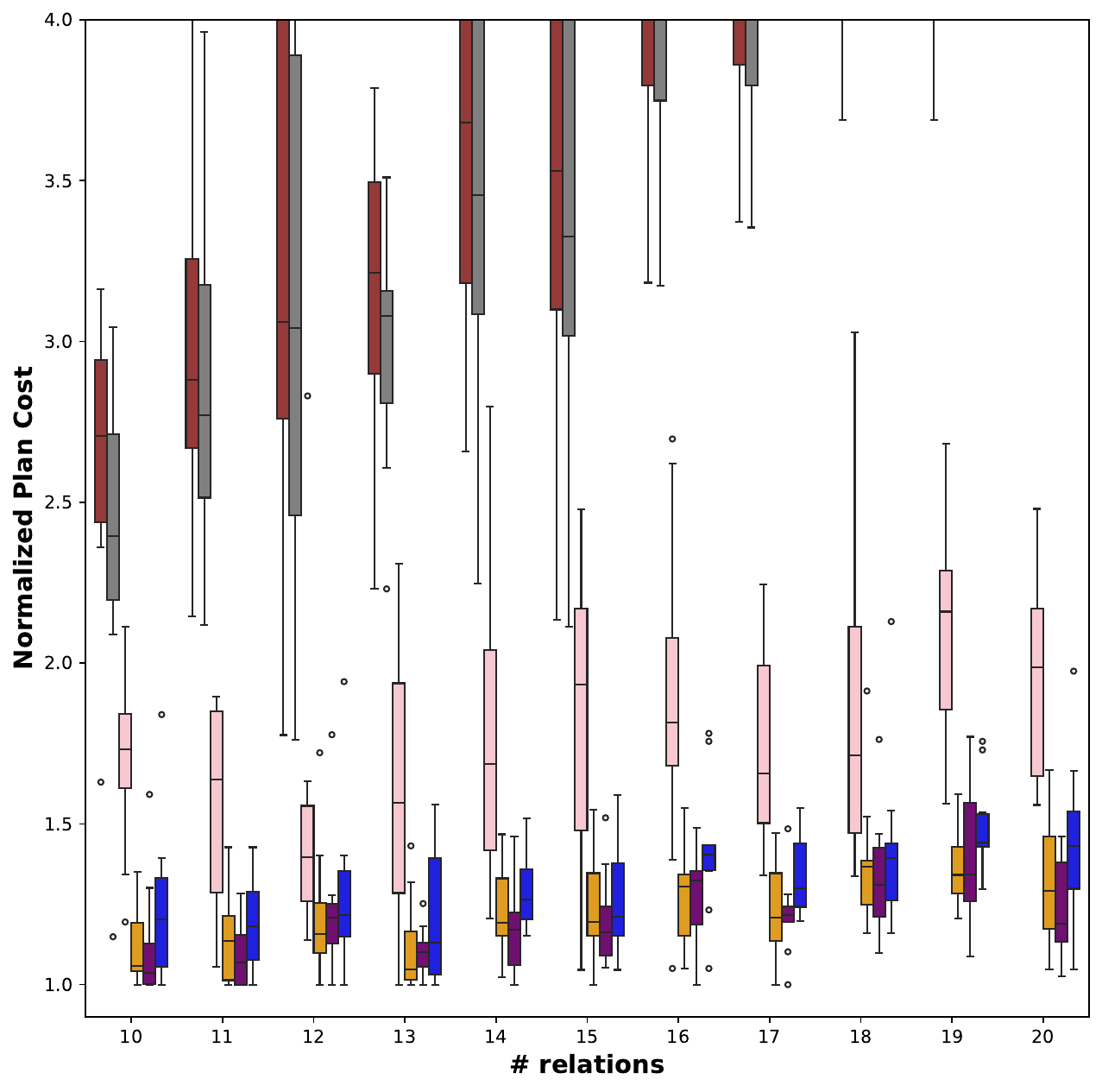}
		\caption{Star}
		\label{fig:top_cout_star}
	\end{subfigure}
	\begin{subfigure}[t]{0.48\textwidth}
		\includegraphics[width=\textwidth]{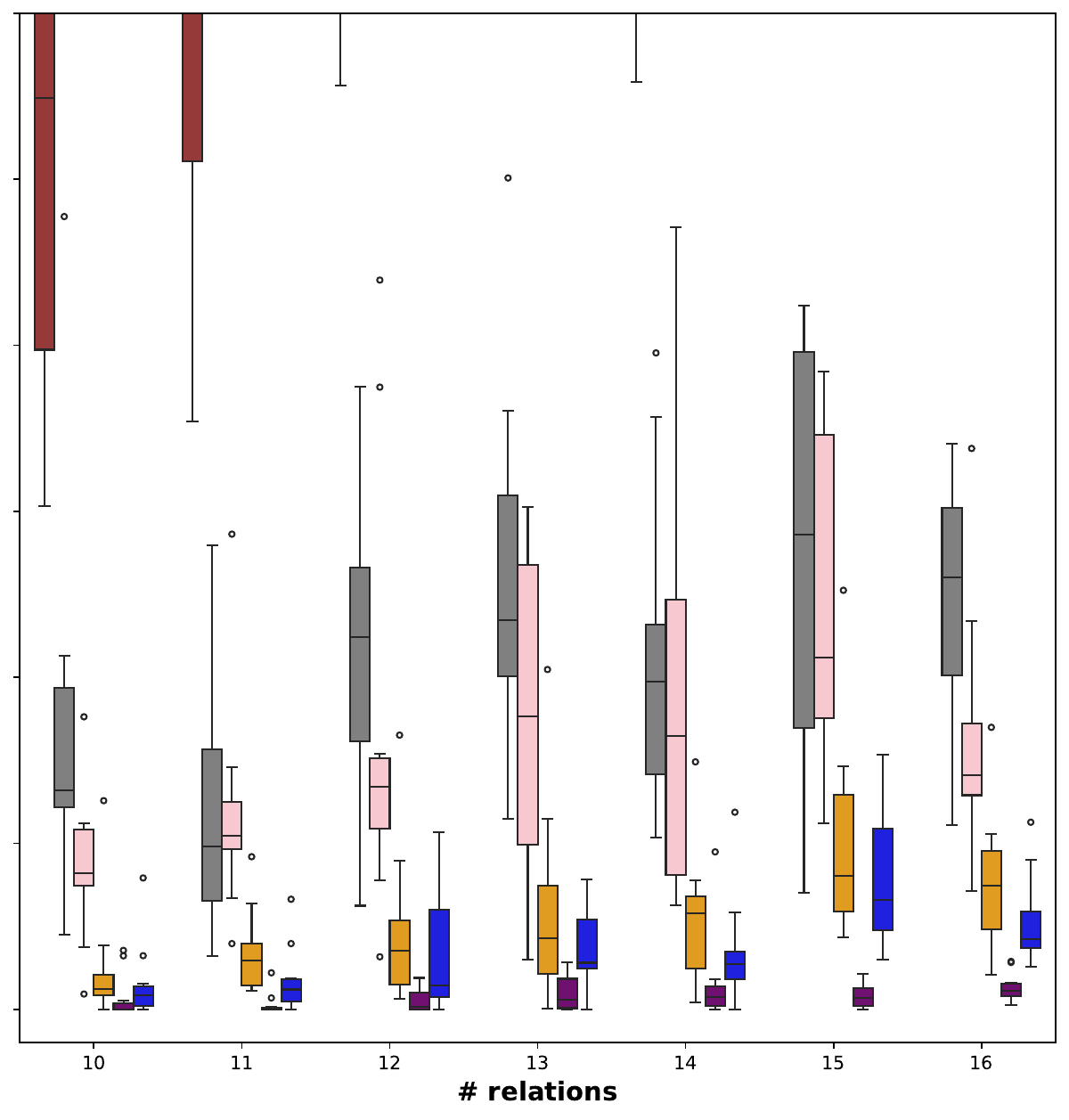}
		\caption{Clique}
		\label{fig:top_cout_clique}
	\end{subfigure}
  	\caption{Cost ratio of selected and optimal query plans. Queries include four graph topologies -- chain, cycle, star, and clique.}
  	\label{fig:top_cout_costs}
\end{figure*}

\begin{figure*}[htbp]
	\centering
	\begin{subfigure}[t]{0.51\textwidth}
		\includegraphics[width=\textwidth]{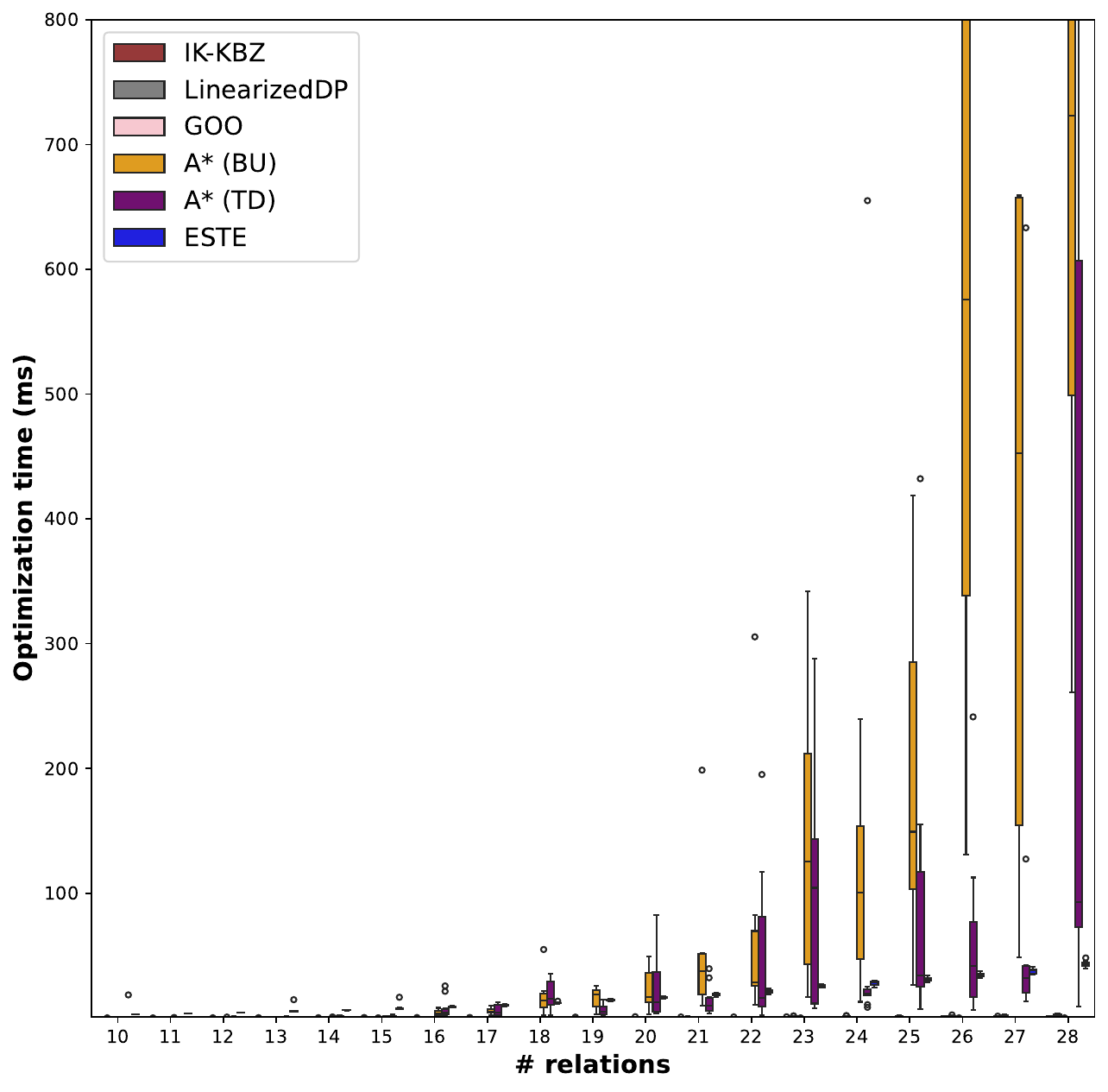}
		\caption{Chain}
	  	\label{fig:top_opt_chain}
	\end{subfigure}
	\begin{subfigure}[t]{0.48\textwidth}
	  	\includegraphics[width=\textwidth]{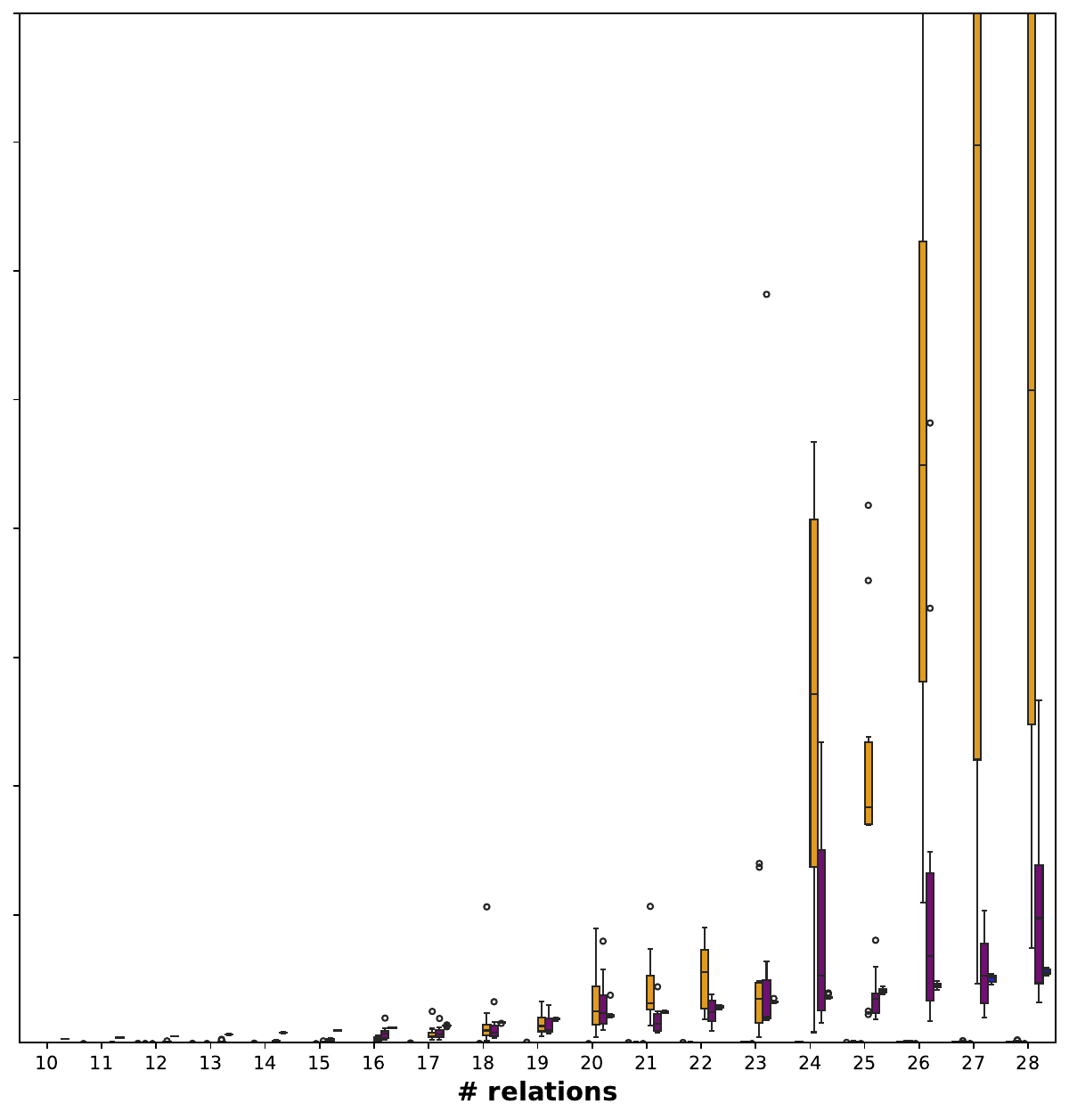}
	  	\caption{Cycle}
	  	\label{fig:top_opt_cycle}
  	\end{subfigure}
	\hfill
	\begin{subfigure}[t]{0.51\textwidth}
		\includegraphics[width=\textwidth]{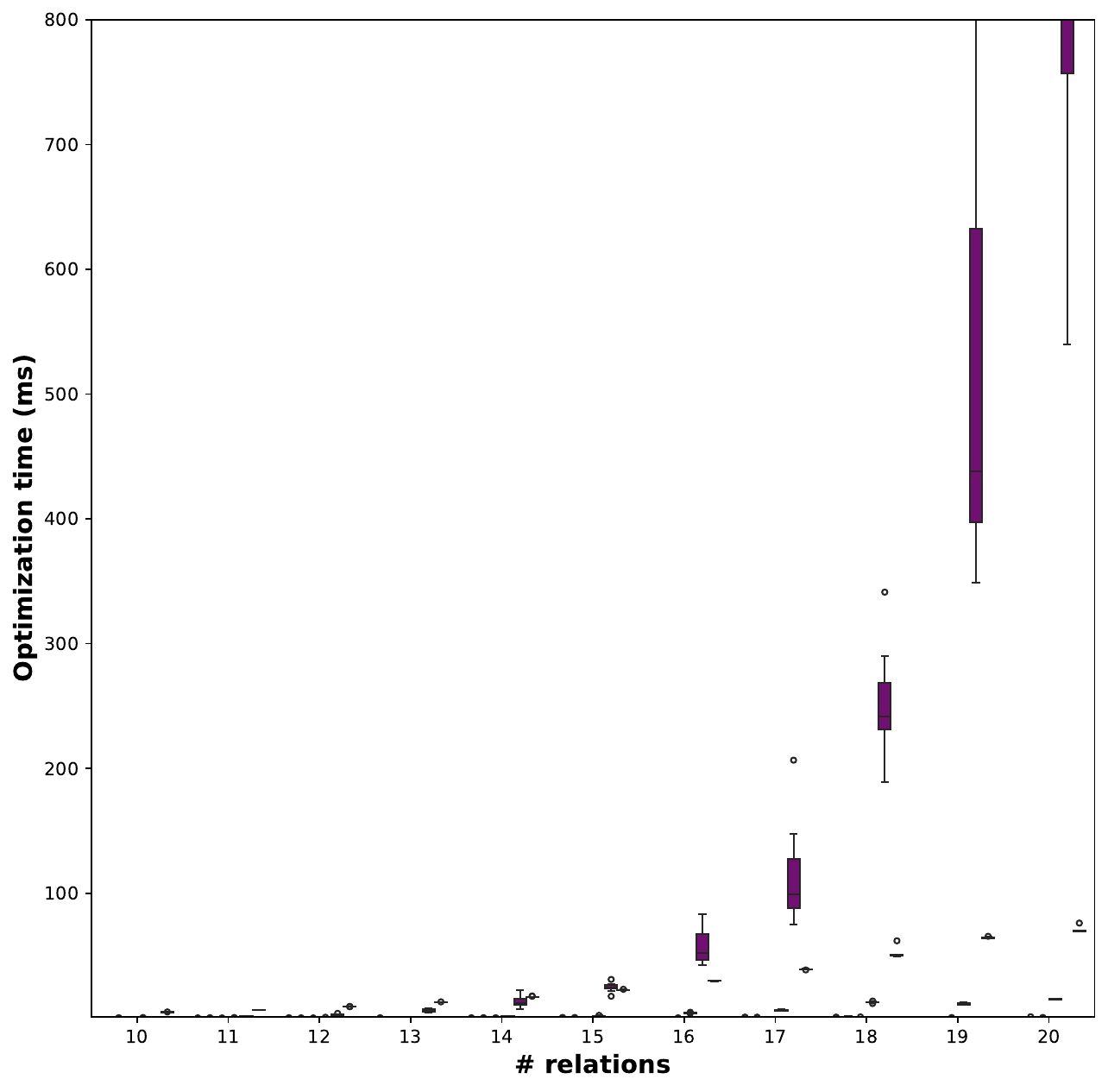}
		\caption{Star}
		\label{fig:top_opt_star}
	\end{subfigure}
	\begin{subfigure}[t]{0.48\textwidth}
		\includegraphics[width=\textwidth]{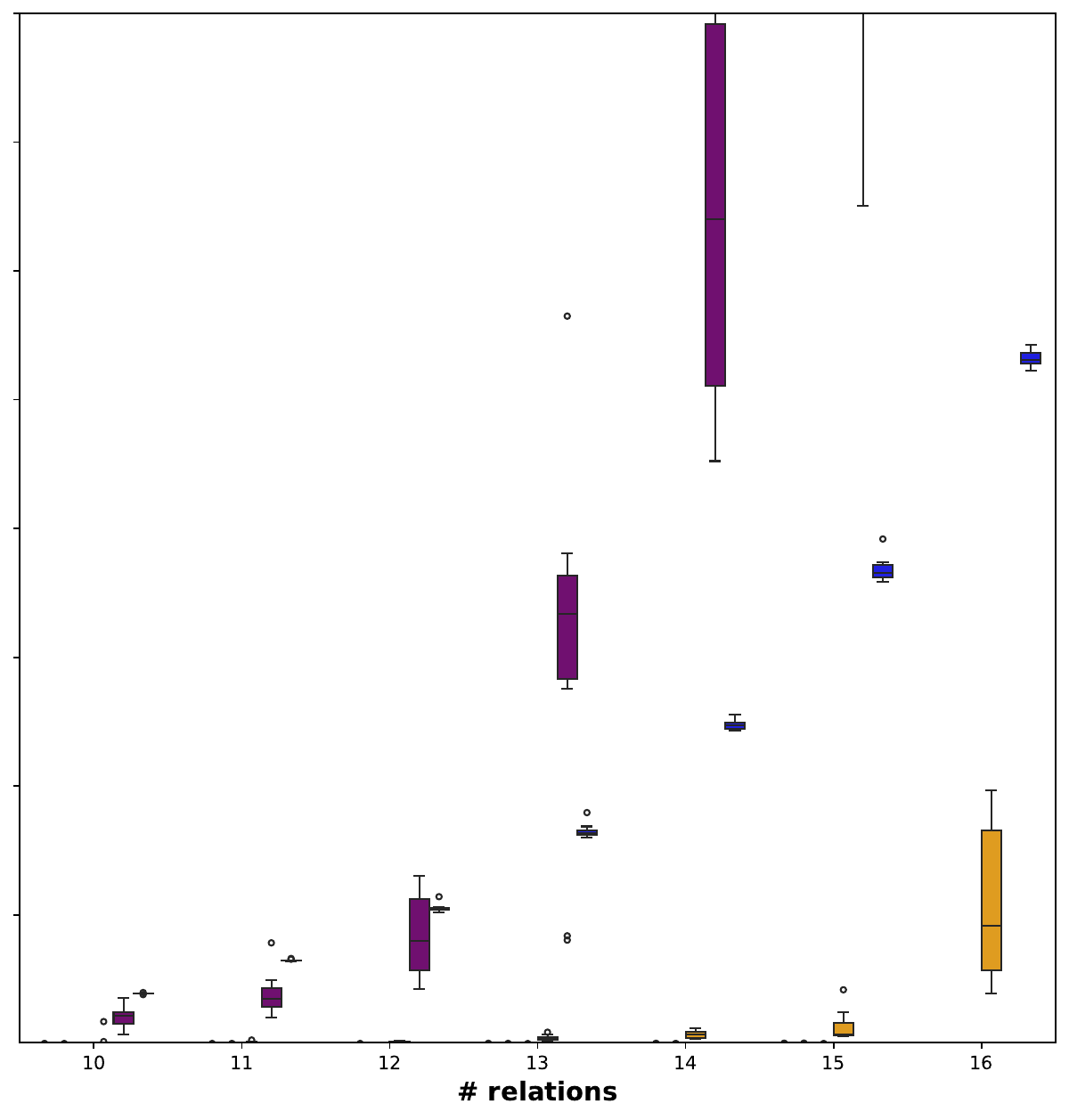}
		\caption{Clique}
		\label{fig:top_opt_clique}
	\end{subfigure}
  	\caption{Optimization time (ms) of selected query plans.}
  	\label{fig:top_cout_opt}
\end{figure*}

\subsection{Evaluation on join graph topologies.}
\textbf{Query plan costs.}
In Figure~\ref{fig:top_cout_costs}, we evaluate the plan costs of heuristic enumeration algorithms over four join graph topologies including chain, cycle, star, and clique. The enumeration algorithms are IK-KBZ, LinearizedDP, GOO, $A^*$ (BU) bottom-up, $A^*$ (TD) top-down, and ESTE. The $C_{out}$ costs of selected plans are the sum of intermediate tables -- shown on the y-axis and normalized to optimal plan costs replicated as in ~\cite{mutable-github,Haffner:HEU:sigmod-2023}. The x-axis shows the complexity of queries in the number of tables. In all topologies, IK-KBZ performs the worst. It is surprising to observe IK-KBZ performing poorly in non-cyclic queries such as chain and star topologies. LinearizedDP and GOO perform similarly except in star queries. These two enumeration algorithms perform worse than the remaining methods -- $A^*$ and ESTE. $A^*$ (BU) and ESTE show the second-best results while $A^*$ (TD) shows a clear dominance. However, $A^*$ methods pose significantly high optimization time, due to operating on its plan graph representation.

\textbf{Optimization time.}
In Figure~\ref{fig:top_cout_opt}, we measure the optimization time of the heuristic enumeration algorithms over four join graph topologies -- shown on the y-axis. The x-axis shows the complexity of queries in the number of tables. Expectedly, IK-KBZ, LinearizedDP, and GOO exhibit the fastest optimization time operation on a join graph. In chain and cycle queries, $A^*$ methods significantly high optimization time and high variance. In star and clique queries, $A^*$ (TD) exhibits a similar trend. ESTE shows linearly increasing optimization time with low variance across join graph topologies.

\subsection{Summary}
The experimental results can be summarized as follows:

\begin{itemize}[leftmargin=*,noitemsep,nolistsep,topsep=0pt]
	\item ESTE leverages a combination of fast-spanning-tree-based heuristics to find the balance between plan quality and computational constraints, especially in large queries. ESTE brings consistency and competitive performance sacrificing minimal extra time in optimization. In large queries, the results suggest that ESTE can be a better alternative than enumeration methods with high optimization time including exhaustive enumeration.
	\item In terms of query plan cost, optimization and execution time, ESTE shows consistent or linearly increasing performance with low variance across different join graph topologies.
\end{itemize}
\section{RELATED WORK}\label{sec:related_work}
A simple and common way to represent a query iss through an undirected join graph. Once constructing the join graph, the objective is to evaluate different join orders, each can be illustrated by a join tree. It is straightforward to see that a join order is a spanning tree connecting all the vertices. To our knowledge, the concept of finding a join order by finding a spanning tree was first introduced in ~\cite{Ibaraki:IK:tods-1984,Krishnamurthy:KBZ:vldb-1986}. The IK-KBZ algorithm, proposed by Krishnamurthy et al.~\cite{Krishnamurthy:KBZ:vldb-1986}, is designed to find an optimal left-deep tree in an acyclic graph. The algorithm is well-suited for star queries, without resorting to cross-joins. As a heuristic extension to handle cyclic queries, the authors suggest finding a spanning tree with the minimum cost, where the cost is calculated as the multiplication of pre-defined selectivities of the edges. Then, the spanning tree is used as an input tree for the IK-KBZ algorithm~\cite{Krishnamurthy:KBZ:vldb-1986}. Fegaras introduces the Greedy Operator Ordering (GOO) algorithm to greedily enumerate a single plan, which can either be a bushy or left-deep tree, on a join graph~\cite{Fegaras:NEWHEU:dexa-1998}. In the join graph, the weights on vertices and edges are cardinalities and selectivities, respectively. At each step, GOO joins two vertices with the minimum cost into a new vertex with updated cardinality weight. The cost is the cardinalities of both vertices multiplied by their selectivity without considering the cost of physical operators. Additionally, if the two vertices share a common adjacent vertex, both edges are merged into one and the weight is updated as the multiplication of both selectivities. The selectivity weights on the other edges remain the same.

The concept of initiating the plan search from each vertex in the join graph, as mentioned in ~\cite{Krishnamurthy:KBZ:vldb-1986,Moerkotte:BQO-book:2023}, assesses more than a single join order, leading to potentially more efficient query plans. Moerkotte proposes the GreedyJoinOrdering-3 to greedily select the next table to join, while considering cross-joins~\cite{Moerkotte:BQO-book:2023}. The cost of joining tables is calculated with respect to the sequence of already joined tables thus maintaining a sorted list of weights does not offer an advantage. This is because the cost of a join operation is dynamic and depends on the previously formed joins, making static weights less relevant.

There are other ways of representing queries. Lee et al.~\cite{Lee:MVP:tkde-2001} defined the join graph where each vertex is a join operator and an edge is present between two vertices if both join operators share a common base table. The authors propose the Maximum Value Precedence (MVP) algorithm to find a directed spanning tree. However, a notable aspect of their join graph definition is that it increases the search space due to the inclusion of spanning trees that feature extra, unnecessary joins. These are referred to as ineffective spanning trees. Finding the shortest path on a directed plan graph was introduced in ~\cite{Negi:FLOWLOSS:pvldb-2021,Haffner:HEU:sigmod-2023}. The plan graph is a directed acyclic graph where a vertex represents a set of tables (subgraph) and an edge is established if one vertex is a subgraph of another. Due to different combinations of table sets, a plan graph typically has a larger number of vertices and edges compared to a join graph.
\section{CONCLUSIONS}\label{sec:conclusion}
In this work, we approach query optimization through the lens of spanning trees, adjusting our aim towards identifying an ordered set of edges that encompasses all vertices within the join graph. The goal is to minimize the cumulative edge costs, which dynamically change as tables are progressively joined. We have tailored Prim's and Kruskal's spanning tree algorithms to our objective, incorporating them into ESTE. Capitalizing on the polynomial-time complexity of spanning tree algorithms, ESTE is proficient at systematically evaluating extensive and diverse segments of the search space. This translates to better consistency in query plan quality and optimization time devoting minimal extra time to optimization. We believe that leveraging spanning tree algorithms extensively in query optimization can be a promising approach to explore even more distinct portions of the search space thus increasing the robustness of query optimizers.

\paragraph*{Acknowledgments.}
This work is supported by NSF award number 2008815.

\bibliographystyle{plain}
\bibliography{references.bib}  

\end{document}